\documentclass[12pt]{article} 

\usepackage{geometry} 
\usepackage{graphicx} 
\usepackage{subcaption} 

\def\eop{\hfill$\fbox{}$\medskip}

\newtheorem{Theorem}{Theorem}

\newtheorem{cor}{Corollary}

\long\def\comment#1\endcomment{}
\date{}

\title{On Preemptive Scheduling of Unrelated Machines Using Linear Programming}
\author{Nodari Vakhania}
\date{}
\begin{document}
\maketitle

\begin{abstract}
We consider a basic problem of preemptive scheduling of $n$ 
non-simultaneously released jobs on a group of $m$ unrelated parallel 
machines so as to minimize maximum job completion time, the makespan. 
In the scheduling literature, the problem is commonly
considered to be solvable in polynomial time by linear programming (LP) 
techniques proposed in Lawler and Labetoulle \cite{ll78}. 
The authors in \cite{ll78} give a LP formulation of the version with 
simultaneously released jobs and show how an optimal solution to this
LP can be used to construct an optimal schedule to the latter problem. 
They also give a linear programming formulation of a related problem for 
non-simultaneously released jobs with the objective to minimize the maximum
job lateness, and show how an optimal solution to this  LP problem 
can be used to construct an optimal solution for non-simultaneously released 
jobs to minimize the maximum job lateness. As the current study shows, for
non-simultaneously released jobs, unlikely, there exist a linear program 
such that a  schedule with the minimum makespan can be constructed based 
on an optimal LP solution. We also prove that, in case no splitting of the 
same job on a machine is allowed (i.e., job part assigned to a machine is 
to be processed without an interruption on that machine), the problem is 
NP-hard. As a side result,  we obtain that, whenever job splitting is not 
allowed, given an optimal LP solution, it is NP-hard to find an optimal 
schedule with the minimum makespan that agrees with that LP solution.  As 
another side result, we obtain that it is NP-hard to find an optimal schedule 
with at most $m-1$ preemptions if jobs are released non-simultaneously. We 
also present two positive results. First, we construct an optimal schedule 
in a low degree polynomial time in case an optimal solution to a modified 
LP formulation that we propose already possesses ``enough'' amount of 
integer components so that it yields no more than 
$m$ preempted jobs parts. We also propose a stronger mixed integer linear 
program formulation. Finally, we extend the schedule construction procedure 
based on an optimal LP solution from Lawler and Labetoulle \cite{ll78} for 
non-simultaneously released jobs.  The extended procedure, among all feasible 
schedules that agree with any feasible (not necessarily optimal) LP solution, generates
one with the minimum makespan. Such procedure is helpful, in particular, because, 
as we show, there may exist no optimal schedule that agrees with an optimal LP
solution if jobs are non-simultaneously released. 
\end{abstract}


\section{Introduction} 

Scheduling unrelated machines to minimize the maximum job completion 
time is a well-known optimization problem. In a group of {\em unrelated} 
machines, a machine $i$ has no universal speed characteristic (machine speed 
is job dependent), in contrast to a group of {\em uniform} machines,  where 
each machine is characterized by a universal speed that extends to a whole
set of jobs. A group of machines with the same speed is commonly referred
to as a group of {\em identical machines} (the speed of every machine is
the same for every job). 
In the scheduling problem that we consider here, commonly 
abbreviated as  $R |r_j; pmtn|C_{\max}$, there are $n$ jobs to be 
performed by $m$ parallel unrelated machines. Job $j$ becomes available 
at its (integer) {\it release time} $r_j$ and it requires an (integer) 
{\it processing time} $p_{ij}$ on machine $i$, $i=1,\dots,m$ (for the sake 
of simplicity, we will refer to jobs and machines by their corresponding 
indexes). A job, 
being processed by a machine, can be interrupted and resumed later on
the same or on any other machine. In a {\it feasible schedule} a machine 
can process at most one job at a time and a job can be processed by at most 
one machine at a time, every job $j$ is assigned to a machine no earlier
than at time $r_j$ and it is completely processed, i.e., 
$\sum_{i=1}^m t_{ij}/p_{ij}=1$, where $t_{ij}$ is the total amount of time 
that machine $i$ spends on job $j$ in that schedule. The objective is to find an 
{\em optimal schedule}, a feasible one in which the  maximum job (machine) 
completion time $C_{\max}$ is the minimum possible. 

In the scheduling literature, this classical scheduling problem is commonly
considered to be solvable in polynomial time by linear programming 
technique from Lawler and Labetoulle \cite{ll78} (we refer the
reader to publicly accessible sources \cite{zoo 1} and \cite{zoo}). 
Lawler and Labetoulle \cite{ll78} proposed a linear programming formulation 
for the version of the above described problem without job release times, 
and have adopted an open shop scheduling method of Gonzalez and Sahni 
\cite{GSopen} for the construction of an optimal schedule for that version 
of the problem based on an optimal LP solution.
A feasible LP solution determines which part of each job is to be
processed by each machine, i.e., it {\em distributes} job parts to machines
without specifying particular starting time of job part(s) on the
corresponding machine(s). Hence, a scheduling stage identifying start time 
of every job part on the corresponding machine is required to transform 
such distribution to a feasible schedule. 

It is commonly accepted  that scheduling problems with
non-simultaneously released jobs are considerably more difficult to solve
than the corresponding versions with simultaneously released jobs (it is
a typical scenario that a polynomially solvable scheduling problem with 
simultaneously released jobs becomes NP-hard if jobs are
released non-simultaneously at arbitrary times). A distribution to linear program
considered in Lawler and Labetoulle \cite{ll78} ``implicitly assumes'' 
that job parts can be assigned to the machines at any time moment. This does 
not apply in case the jobs are non-simultaneously released. Such a ``lack of 
information'' essentially complicates the use an optimal fractional LP solution 
for the construction of an optimal preemptive schedule for non-simultaneously
released jobs. We suggest two alternative linear programs for problem 
$R |r_j; pmtn |C_{\max}$ that are more ``flexible'' since they take somehow 
into account job release times. The second LP, is however, a mixed integer 
linear program. Even the latter LP is not strong enough in the sense that
an optimal schedule not necessarily agrees with an optimal distribution 
generated for that LP. In fact, there may exist no optimal schedule that
agrees with any optimal distribution to any of the linear programs that we
consider here. Moreover, as we argue, unlikely, there exists a ``strong 
enough'' linear program such that an optimal solution to that linear program 
can be used for the construction of an optimal schedule to problem 
$R |r_j; pmtn|C_{\max}$ with non-simultaneously released jobs. If however, 
an optimal solution to a modified LP formulation that we propose, possesses ``enough'' 
amount of integer components (so that it yields no more than $m$ preempted jobs 
parts), then an optimal schedule can be constructed in a low degree 
polynomial time based on this optimal LP solution, as we show in Section 4. 
We also extend the schedule construction procedure 
based on an optimal LP solution of Lawler and Labetoulle \cite{ll78} for 
non-simultaneously released jobs.  The extended procedure, among all feasible 
schedules that agree with any feasible (not necessarily optimal) LP solution, generates
one with the minimum makespan. The extended procedure is helpful, in particular, 
because any  optimal schedule may only agree  with a non-optimal LP
solution if jobs are non-simultaneously released. As this study shows, for
non-simultaneously released jobs, unlikely, there exist a linear program 
such that a  schedule with the minimum makespan can be constructed based 
on an optimal solution to that linear program. 

We also give NP-hardness results. In particular, 
if no splitting of the same job on a machine is allowed (i.e., job part assigned 
to a machine is to be processed without an interruption on that machine), then we 
show that the problem with non-simultaneously released jobs is NP-hard. As a side 
result,  we obtain that, whenever job splitting is not allowed, given an optimal 
LP solution, it is NP-hard to find an optimal schedule with the minimum makespan 
that agrees with that LP solution. As another side result, we obtain that it is 
NP-hard to find an optimal preemptive schedule with at most $m-1$ preemptions. 
This result somehow extends an earlier known fact that scheduling identical
machines with at most $m-2$ preemptions, even for simultaneously released jobs,
is NP-hard \cite{geo}. 

We complete this section by a very brief look at a few earlier known related 
results. Non-preemptive scheduling on identical machines of simultaneously released 
jobs $P ||C_{\max}$ is already NP-hard. This is in contrast to the preemptive 
case, even on a group of uniform machines $Q |pmtn |C_{\max}$, for which a
low-degree polynomial time algorithm is known Gonzalez and Sahni 
\cite{GS}. The setting with non-simultaneously released jobs 
$Q |r_j,pmtn |C_{\max}$ remains polynomially solvable Labetoulle et al. 
\cite{lllr}, as well as the version with unrelated machines but simultaneously
released jobs $R |pmtn |C_{\max}$ Lawler and Labetoulle \cite{ll78}. As already noted, 
the authors in \cite{ll78} give a linear programming formulation of the problem
and show how an optimal solution to this LP can be used to construct an optimal 
solution to the latter problem. Lawler and Labetoulle \cite{ll78} also give a linear
programming formulation of a related problem $R | pmtn |L_{\max}$ where the 
objective is to minimize the maximum job {\it lateness} $L_{\max}$ (the lateness 
$L_j$ of job $j$ in schedule $S$ is the difference between the completion time of 
that job in that schedule and its {\it due-date} $d_j$, a ``desirable'' completion 
time for that job, and $L_{\max}$ is the maximum job lateness). This formulation
allows to use their schedule construction procedure to generate an optimal schedule 
to that problem.

\section{Alternative linear programs}

In this section we present linear programs that have been used for 
scheduling unrelated machines. 
The following linear program LP1$(C_{\max })$ was successfully used for 
an approximate solution of the non-preemptive version of the problem
$R ||C_{\max}$ with simultaneously released jobs first by Potts \cite{Potts} 
and later in \cite{round}:

Minimize $C_{\max }$ \\
Subject to\\
$\sum\limits_{j=1}^n x_{ij}p_{ij}\le C_{\max}, \ \ i=1,\dots ,m$\\
$\sum\limits_{i=1}^m {x_{ij}}=1,\ \ j=1,...,n$\\
$x_{ij}\ge 0, \ \ \ i=1,...,m, \ \ j=1,\dots ,n.$\\

In this linear program entry $x_{ij}= t_{ij}/p_{ij}$ represents the part of  
job $j$ to be processed by machine $i$, for  $j=1,\dots,n$ and $i=1,\dots,m$. 
These entries define the corresponding 
{\it distribution} of job parts on machines. 
Unlike a  schedule which is a mapping that assigns to 
each job specific time interval(s) on one or more machines, a 
distribution, instead of these time intervals, defines only their lengths 
on the corresponding machines. Because of real assignment variables, a distribution
may split a job in different parts and assign these parts to different machines.
Note that a distribution involves no start times and only assigns job parts to 
machines (hence, there is an infinite number of feasible schedules respecting
a given distribution). In particular, a solution to a linear program is a distribution 
that explicitly indicates which fraction of each job is assigned to each machine.
We refer the reader to \cite{round} for related formal definitions,  concepts 
and properties. 

A  distribution to linear program $LP1(C_{\max })$
has a nice property that it possesses a large amount of integer (zero) entries 
so that it yields  at most $m-1$  preempted jobs. Then such distribution can be 
rounded to (an integer) feasible approximate non-preemptive solution to problem 
$R ||C_{\max}$, as suggested by Potts \cite{Potts}, where a complete enumeration 
of at most $m-1$ preempted jobs on $m$ machines is carried out. This results in 
a 2-approximation solution in time, polynomial in $n$ and exponential in $m$. 
Using a modified linear program combined with a binary search, a rounding that 
guarantees a 2-approximation solution in polynomial (in both $n$ and $m$) time
was achieved in Lenstra et al. \cite{Lenstr}. A new method of rounding an optimal
distributionmto linear program $LP1(C_{\max })$ proposed in \cite{round} yielded 
an improvement of the bound 2 to $2-1/m$. This latter bound is the best possible 
that can be obtained by rounding a distribution to a feasible non-preemptive schedule. 

In an optimal distribution to linear program LP1$(C_{\max })$,    
the total length of the parts of a job assigned to the machines can be longer 
than the minimized magnitude $C_{\max }$. Hence, such distribution is inappropriate for 
the preemptive case $R |pmtn|C_{\max}$. Linear program LP2$(C_{\max })$ which bounds 
the length of the assigned parts of each job was studied in Lawler and Labetoulle
\cite{ll78}. This linear program requires an additional set of $n$ restrictions 
(\ref{2}) (one additional restriction for each job):   

Minimize $C_{\max }$ \\
Subject to\\
\begin{equation}\label{1}
\sum\limits_{j=1}^n x_{ij}p_{ij}\le C_{\max}, \ \ i=1,\dots ,m
\end{equation}
\begin{equation}\label{2}
\sum\limits_{i=1}^m x_{ij}p_{ij}\le C_{\max}, \ \ j=1,\dots ,n
\end{equation}
$\sum\limits_{i=1}^m {x_{ij}}=1,\ \ j=1,...,n$\\
$x_{ij}\ge 0, \ \ \ i=1,...,m, \ \ j=1,\dots ,n.$\\

Because of these additional $n$ restrictions (\ref{2}), the total number of 
basic (non-zero) variables is no more bounded by $m-1$ (in case an optimal 
distribution to linear program LP2$(C_{\max })$ still yields no more than $m-1$ 
preemptions, it can easily be transformed to an optimal preemptive schedule
with at most $m-1$ preemptions, see Section 4). 
If an optimal distribution to linear program $LP2(C_{\max })$ yields
more than $m$ preempted job parts, an optimal schedule to problem 
$R |pmtn|C_{\max}$ can still be constructed in polynomial time. Lawler and 
Labetoulle \cite{ll78} adopted open shop scheduling technique from 
Gonzalez and Sahni \cite{GSopen} for constructing an optimal feasible
schedule from an optimal distribution to linear program LP2$(C_{\max })$ (note 
that an open shop instance can be already seen as a distribution). We describe
this method in more detail in Section 4. 

Suppose we have a feasible schedule for an instance of problem $R |pmtn|C_{\max}$  
respecting  an optimal distribution $\{x_{ij}\}$ to linear program LP2$(C_{\max })$
with the makespan 

\begin{equation}\label{Cmax}
C_{\max}= \max\{ \max_i\sum\limits_{j=1}^n x_{ij}p_{ij}, 
                  \max_j\sum\limits_{i=1}^m x_{ij}p_{ij}.\}
\end{equation}

\noindent Then this schedule is clearly optimal. Such an optimal schedule is 
constructed in Lawler and Labetoulle \cite{ll78}. 

During the construction of the schedule in \cite{ll78}, job parts can be assigned 
to machines at any time moment. Such an approach is not particularly useful if
the jobs are not simultaneously released. Such ``lack of required restrictions''
questions the usefulness of an optimal fractional 
solution (an optimal distribution) to a linear program for the construction of an 
optimal preemptive schedule for  problem $R |r_j; pmtn |C_{\max}$. In particular, 
bound (\ref{Cmax}) is no more attainable, i.e., for a given instance of 
problem $R |r_j,pmtn|C_{\max}$, there may exist no feasible schedule with 
makespan $C_{\max }$ respecting a given optimal distribution to linear program 
LP2$(C_{\max })$ (e.g., consider an optimal distribution in which the sum of the 
job parts assigned to some machine is  $C_{\max}$ but no job assigned to that
machine is released at time 0). Furthermore, no feasible schedule 
respecting that optimal distribution might be optimal. This assertion is true 
for linear program LP2$(C_{\max })$, and is also true for more enhanced linear 
programs that we give later on.

\section{NP-hardness results}

Let $R |r_j, pmtn \ - \ no split |C_{\max}$ denote the version of problem 
$R |r_j, pmtn |C_{\max}$ in which no job part assigned to a machine can be
split on that machine (it is to be processed continuously on that machine). 
No splitting is a reasonable assumption in a number of applications where it 
is undesirable to interrupt a currently running job on a machine in favor 
of another job (such split would cause unreasonable amount of 
machine reset and setup times). 

\begin{Theorem}\label{1}
$R |r_j, pmtn \ - \ no split |C_{\max}$ is NP-hard.
\end{Theorem} 
Proof. We show that the decision version of $R |r_j, pmtn \ - \ no split |C_{\max}$
is NP-complete using the reduction from an NP-complete PARTITION problem. Consider 
an arbitrary instance of this problem with $k$ items $\{z_1,\dots, z_k\}$  
and with $M = \sum_{i=1}^k z_i$. Let $P_1$ and $P_2$  be a partition of the 
$k$ items with $\sum_{l\in P_1} z_l= \sum_{l\in P_2} z_l = M/2$, a solution 
to PARTITION ($P_1\cup P_2=\{z_1,\dots, z_k\}$, $P_1\cap P_2=\emptyset$). 

Our scheduling instance consists of $4+k$ jobs on three machines, where 
$Z_1,\dots,Z_k$ are {\em partition} jobs. All partition jobs are released at 
time 3. The processing times of these jobs are such that $p_{2Z_j}=2z_j/M$ 
and $p_{iZ_j}=\infty$, for $j=1,\dots,k$ and $i=1,3$ (note that the total 
length of the partition jobs is $2$). The processing times of the remaining 
four jobs are defined as follows. $p_{11}=p_{21}=p_{31}=6$, $p_{13}=2$, 
$p_{24}=3$ and $p_{32}=5$. All the unspecified processing times are infinity
(large enough numbers). Job 3 is released at time 4 and the remaining jobs
are released at time 0 (except the partition ones which are released at time 3).  

As it is easy to see, the processing time 6 of job 1 is a lower bound on the 
optimal schedule  makespan. In a schedule with this makespan (see Fig. 1): \\
On machine 1, job $J^3$ cannot be started earlier than at its release time 4 and 
can be completed at time 6, hence job 1 starts at its release time 0 and competes at 
time 4. On machine 3, job $J^2$ is to occupy 5 time units. Hence, only one time
unit is left where job $J^1$ can be processed on that machine. Therefore, one 
unit of time of job $J^1$ is to be processed on machine 2. All partition jobs are to
be processed also on machine 2. None of the partition jobs can start earlier than 
at time 3 whereas job $J^4$ is to occupy 3 time units on machine 2. Since machine
1 is running job $J^1$ in interval $[0,4)$, job  $J^4$ is needs to occupy the
first 3 time units on machine 2 and is to be followed by the partition jobs. Since 
the assigned to machine 3 part of job $J^2$ cannot be interrupted on that machine, 
the remaining unprocessed unit time of job $J^1$ can only be executed within the 
interval $[4,5)$ on machine 2. Hence, exactly the intervals $[3,4)$ and $[5,6)$ 
are left for the partition jobs. 

\begin{figure}[hbtp]
\centering
\includegraphics[width=2.5cm]{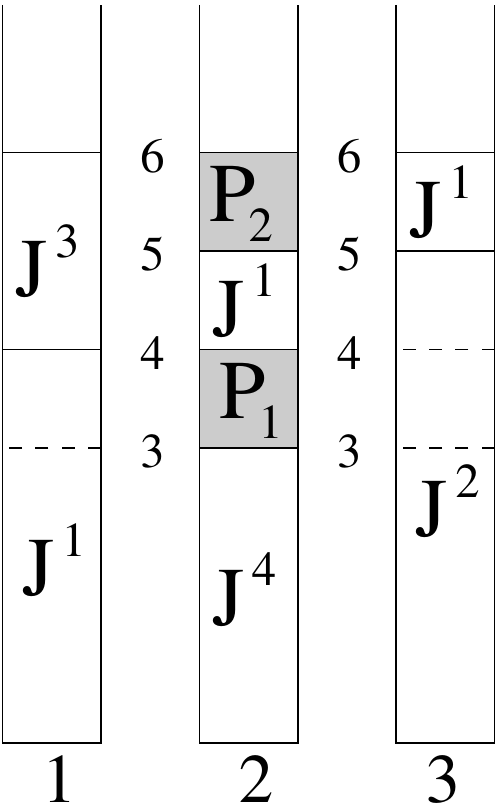}
\caption{An optimal schedule with makespan 6. Dark regions represent partition jobs.}
\label{fig1}
\end{figure}

It should now be apparent that there exist a schedule of length 6 if and only if 
the partition instance has a ``yes'' answer, i.e., there exists a partition of
set $\{z_1,\dots, z_k\}$ into sets $P_1$ and $P_2$ with equal length 1. In other
words,  an optimal schedule with makespan 6 provides a solution to PARTITION, 
and vice-versa, if PARTITION has a solution then the above optimal 
schedule can be constructed in polynomial time.\eop 

\begin{cor}
Given an instance of  problem $R |r_j, pmtn |C_{\max}$, it is NP-hard to find
an optimal solution with at most $m-1$ preemptions. 
\end{cor}
Proof. By Theorem \ref{1}, $R |r_j, pmtn \ - \ no split |C_{\max}$ 
is NP-hard. No split on any machine yields at most $m$ preempted job parts 
(one job part on every machine), hence at most $m-1$ preemptions.\eop

\begin{cor}\label{sch-con}
Given an optimal distribution, it is NP-hard to find a schedule for problem 
$R |r_j, pmtn \ - \ no split |C_{\max}$
with the minimum makespan respecting that distribution.
\end{cor}
Proof. From the proof of Theorem \ref{1}, the PARTITION instance is to be
solved if job $J^2$ is not allowed to be split on machine 3. The corollary
follows as the distribution respected by the schedule of Fig. 1 is optimal.\eop

\section {Scheduling little-preemptive distributions}

In this section we show how an optimal schedule can be constructed in case
an optimal distribution to  linear program LP2$(C_{\max })$ yields at
most $m$ preempted job parts. We first consider the case of simultaneously
released jobs. Then we introduce a new liner program and consider the
case when jobs are not simultaneously released.

\subsection{Simultaneously released jobs}

Suppose an optimal distribution to linear program LP2$(C_{\max })$ 
yields no more than $m-1$ preemptions, i.e., there are at most $m$ 
(preempted) job parts, hence at most one such part on any machine. 
We can convert such distribution to an optimal preemptive schedule
for problem $R |pmtn |C_{\max}$
with at most $2m-4$ preemptions as follows. Sort the preempted jobs in
non-increasing order of their total processing times (the total
processing time of job $j$ in  distribution $\{x_{ij}\}$ is 
$\sum\limits_{i=1}^m x_{ij}p_{ij}$). Take the first job $j$ in the list
and consider its parts in the ascending order of the machine indexes to
which these parts are assigned. Let $i_1<\dots<i_k$ be these indexes (ones 
of the corresponding machines). Schedule the first part of job $j$ at time 
0 on machine $i_1$, schedule the second part at the completion 
time of the first part on machine $i_2$, and so on, schedule the last 
$k$th part at the completion time of the preceding $(k-1)$th part on 
machine $i_k$. Due to inequalities (\ref{2}), the completion time of the 
last scheduled part of job $j$ cannot be larger than $C_{\max }$. Now
take the second job in the list and schedule its parts similarly, and
so on. Since there is at most one preempted job part on any machine, 
no conflicts between preempted parts of different jobs will occur. 
The remaining entire job parts are scheduled in the remained idle time slots 
from time 0 in any order  without creating an idle time in between 
the jobs. In case an overlapping of an entire part with an earlier included 
preempted job part occurs,  the entire job part is preempted and
resumed at the completion time of the latter preempted job part. Due to 
inequalities (\ref{1}), the makespan of the resultant feasible schedule will 
not surpass  $C_{\max }$. Note that no additional preemption will 
occur on any machine on which a preempted job part is scheduled the first
or the last on that machine. In the worst case, there is only one preempted
job distributed among all the machines. Then $m-3$ additional preemptions
will occur, hence the $2m-4$ is an upper bound on the number of preemptions.

\subsection{Non-simultaneously released jobs}

First, let us modify linear program LP2$(C_{\max })$ by replacing 
inequalities (\ref{2}) with the following set of inequalities that 
take into account the release time of each job: 

\begin{equation}\label{2'}
r_j + \sum\limits_{i=1}^m x_{ij}p_{ij}\le C_{\max}, \ \ j=1,\dots ,n. 
\end{equation}

The resultant new linear program LP3$(C_{\max })$ properly reflects 
the desired restriction for each job. 

Suppose now an optimal distribution to linear program LP3$(C_{\max })$ yields
no more than $m-1$ preemptions. The above described schedule construction 
procedure for simultaneously released jobs can easily be extended 
for non-simultaneously released jobs for problem $R |r_j; pmtn |C_{\max}$.
We again consider job parts in ascending order of the corresponding machine
indexes. Now, instead of starting each first (according to this order) part of 
the next job from the list at time 0, it is scheduled at the release time of 
that job. The entire (non-preempted) jobs are scheduled in non-decreasing 
order of the their release times on each machine. Due to the modified 
inequalities (\ref{2'}) (and the fact that there is at most one preempted job
part on any machine), again, any of the preempted jobs will complete no
later than at time $C_{\max }$.

\section{``Weaknesses'' of linear programs} 

For a given feasible schedule to an instance of the scheduling problem 
$R |r_j; pmtn |C_{\max}$ with makespan $C_{\max}$, let 
$\{x_{ij}\}$ be the distribution that respects this schedule. The values 
$C_{\max}$ and $\{x_{ij}\}$ form a feasible solution to linear
programs LP3$(C_{\max })$ and LP2$(C_{\max })$. 
A less obvious question is, given  a  distribution to linear program 
LP3$(C_{\max })$, whether there is an optimal 
schedule respecting that distribution with the makespan $C_{\max}$. If
this assertion is true, then  a feasible schedule respecting
an optimal distribution will be optimal. 

Thus a feasible schedule respecting an optimal distribution is trivially optimal 
if its makespan is $C_{\max }$. Lawler and Labetoulle \cite{ll78} constructed 
schedules respecting optimal distributions to linear program LP2$(C_{\max })$ 
with makespan is $C_{\max }$ for simultaneously released jobs. In contrary to 
the case with simultaneously released jobs, an optimal schedule respecting an 
optimal distribution to linear 
program LP3$(C_{\max })$ (and linear program LP2$(C_{\max })$) not necessarily 
attains makespan $C_{\max }$ if jobs are  non-simultaneously released. 
In particular, inequalities (\ref{1}) no 
more reflect actual restrictions imposed by job release times on the
completion time of each machine, since it may not be possible to schedule
a job assigned to a machine   at the earliest idle-time moment
on that machine. In particular,  restrictions (\ref{1}) are no more necessarily
satisfied in an optimal schedule where the completion time of a machine may 
be larger than $C_{\max }$  (recall an earlier 
mentioned simple scenario where no job assigned to a machine is released at 
time 0). Even restrictions (\ref{2'}) may not be satisfied in an optimal schedule.  

We illustrate these points on small problem instances in the following examples.
For none of these instances the procedure from Section 4.2 can guarantee the 
construction of an optimal schedule with makespan $C_{\max }$. 
Our examples illustrate that in an optimal schedule 
constructed from an optimal distribution to linear program LP3$(C_{\max })$,
neither restrictions (\ref{1}) nor restrictions (\ref{2'}) might be satisfied  
(the completion time of at least one job in that schedule can be larger than 
$C_{\max }$). More importantly, a bit later we will see that an optimal schedule 
not necessarily respects an optimal distribution. 

\smallskip 

{\bf Example 1.} 
Let us consider a problem instance where an optimal distribution assigns just 
two different job parts to the same machine (hence the schedule 
construction procedure from Section 4.2 already cannot be applied). We have 
five jobs $2,\dots,6$ on four machines $1,\dots,4$ such that:\\
$r_3=3,\ r_2=2, \ r_4=8, \ r_5=0, \ r_6=5$.\\
$p_{13}=p_{33}=15$, $p_{22}=p_{32}=p_{42}=16$, and $p_{14}=p_{26}=p_{45}=13$.
The processing times of these jobs on the remaining machines are infinities
(large enough numbers). Note that in any feasible schedule, jobs 4, 6 and
5 can only be processed by machines 1,2 and 4, respectively. Job 2 is to be
distributed among machines 2,3 and 4, and job 3 is to be distributed  on machines 
1 and 3. 


\begin{figure}[ht!]
\centering
\begin{subfigure}[b]{0.4\linewidth}
\centering
\includegraphics[width=0.5\linewidth]{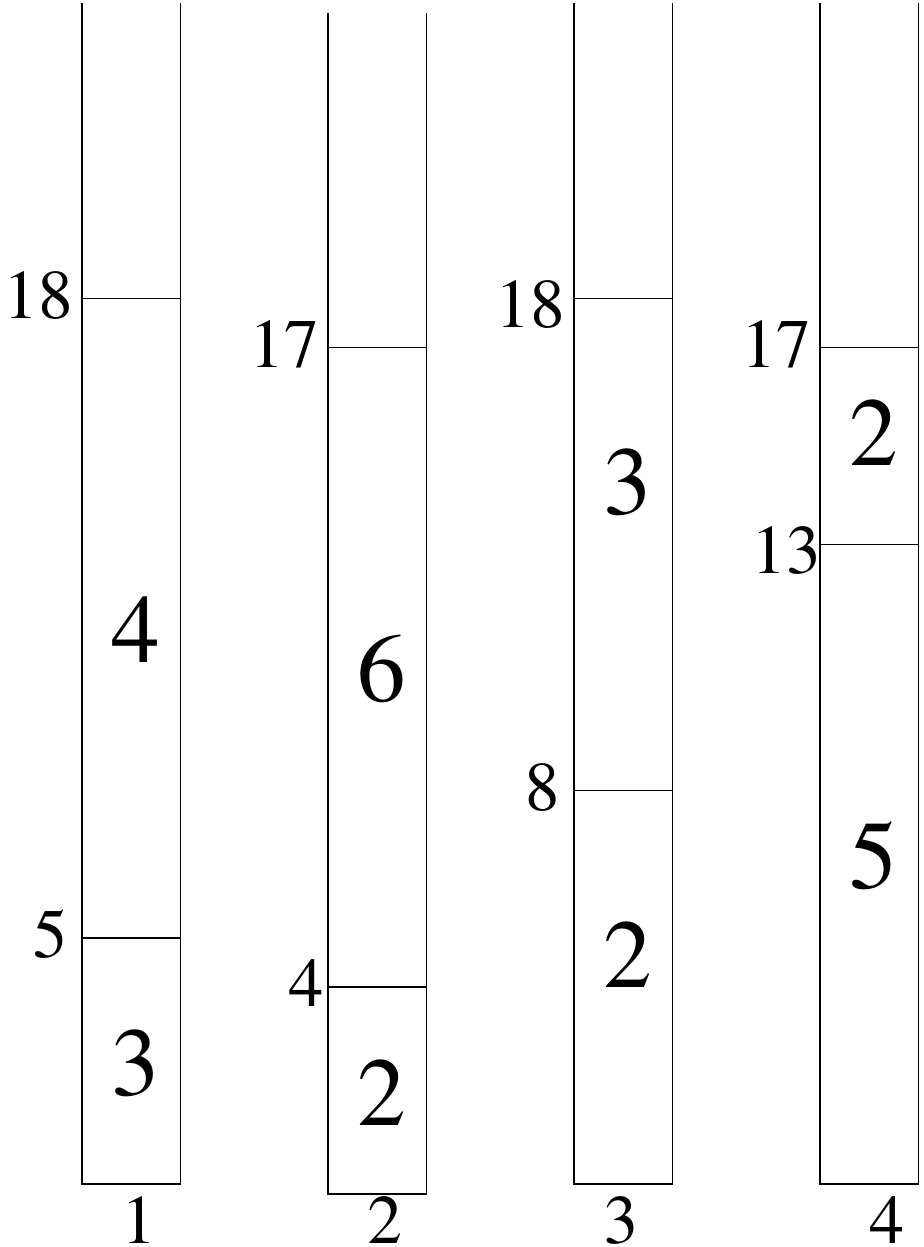}
\caption{A non-feasible schedule respecting distribution 1} 
\label{fig2a}
\end{subfigure}
\hspace{0.1 cm}
\begin{subfigure}[b]{0.4\linewidth}
\centering
\includegraphics[width=0.5\linewidth]{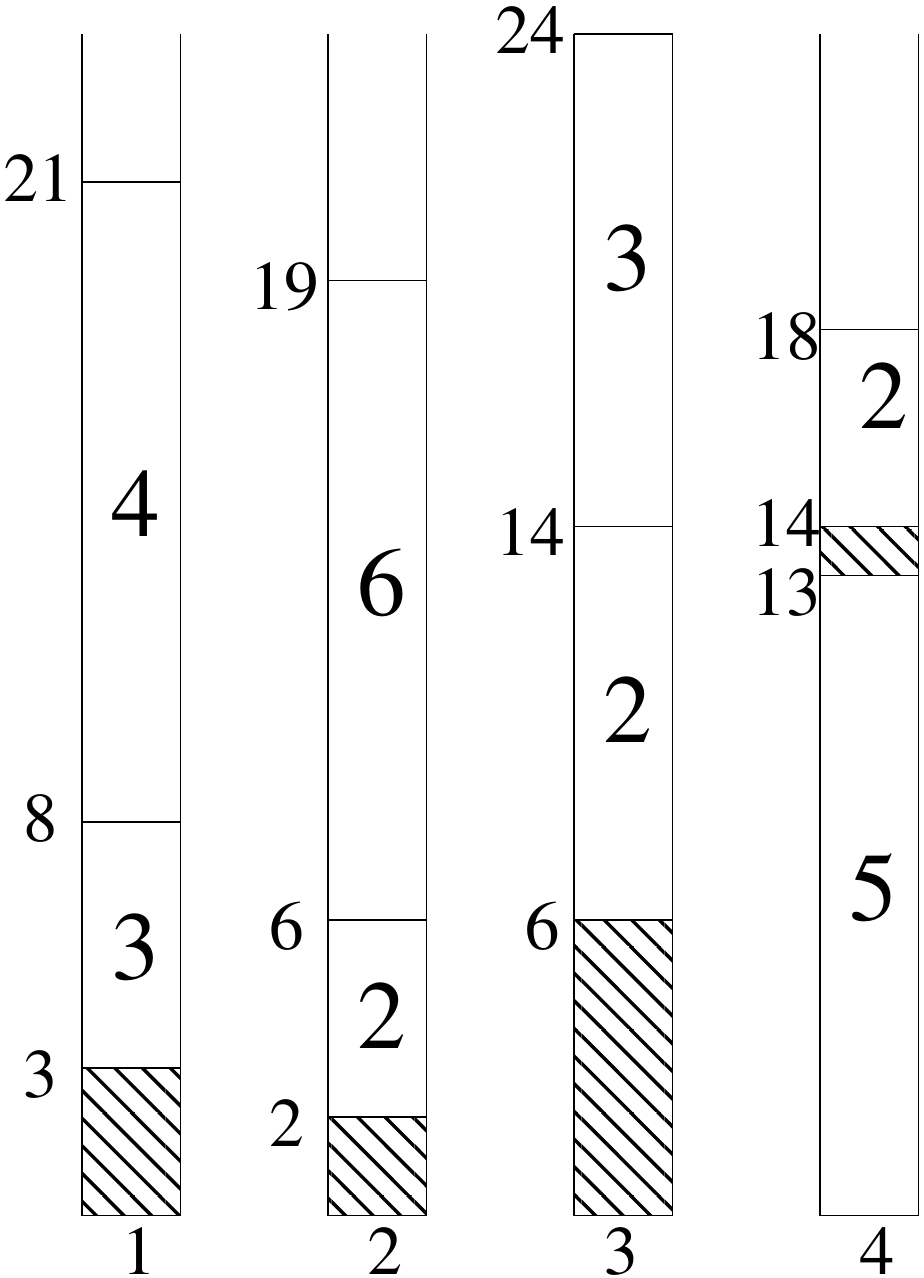}
\caption{A feasible schedule respecting distribution 1}
\label{fig2b}
\end{subfigure}
\hspace{0.1 cm}
\begin{subfigure}[b]{0.4\linewidth}
\centering
\includegraphics[width=0.5\linewidth]{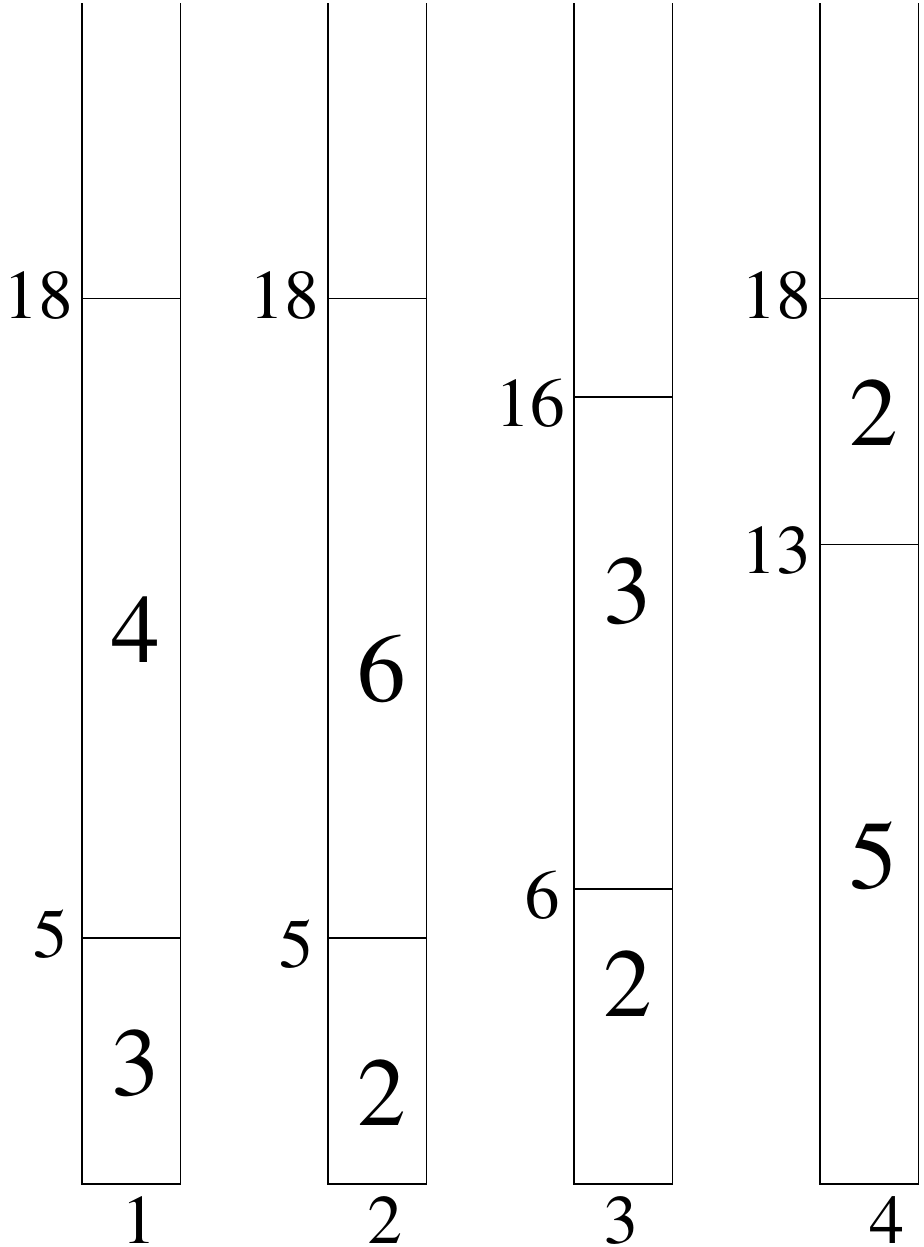}
\caption{A non-feasible schedule respecting distribution 2}
\label{fig2c}
\end{subfigure}
\centering
\begin{subfigure}[b]{0.4\linewidth}
\centering
\includegraphics[width=0.5\linewidth]{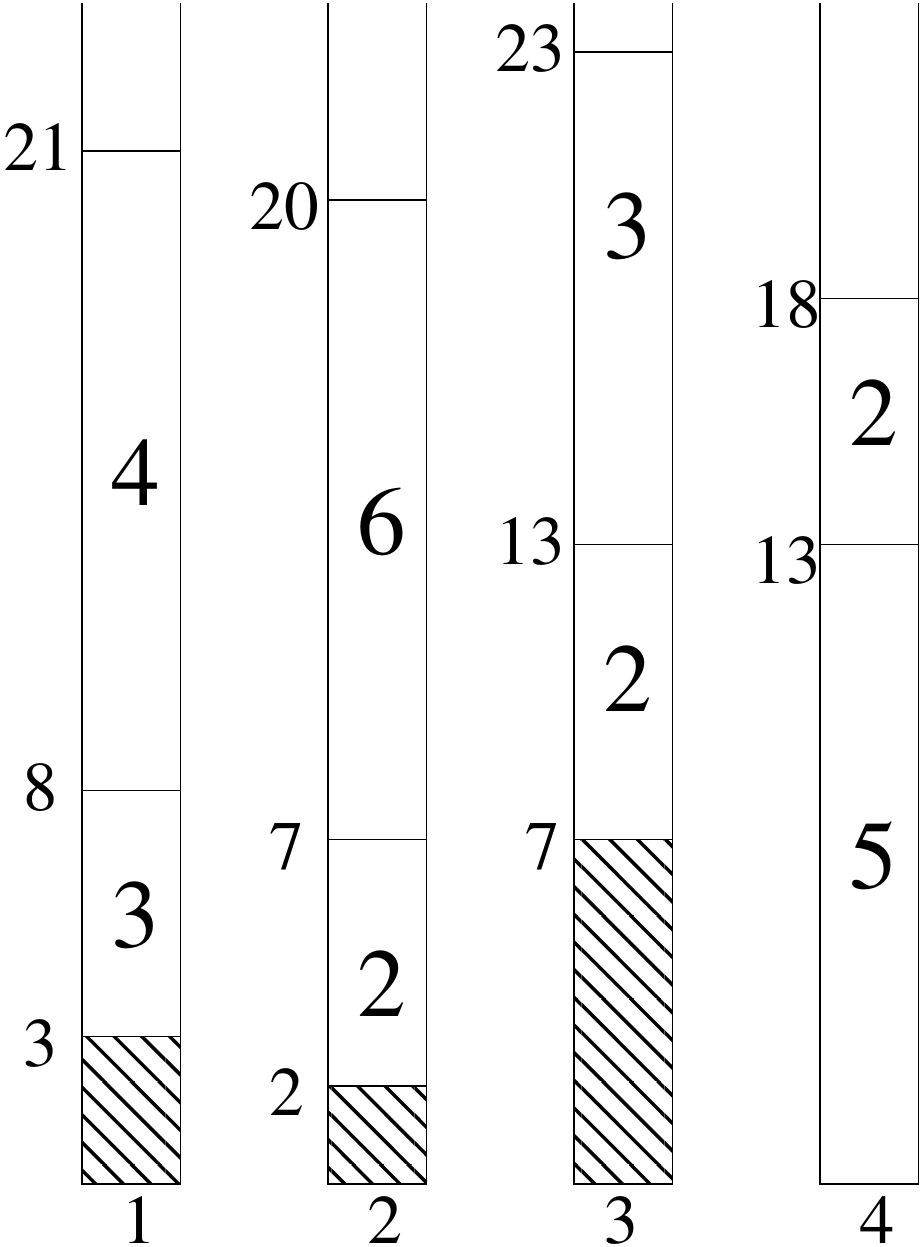}
\caption{A feasible schedule respecting distribution 2} 
\label{fig2d}
\end{subfigure}
\hspace{0.1 cm}
\begin{subfigure}[b]{0.4\linewidth}
\centering
\includegraphics[width=0.5\linewidth]{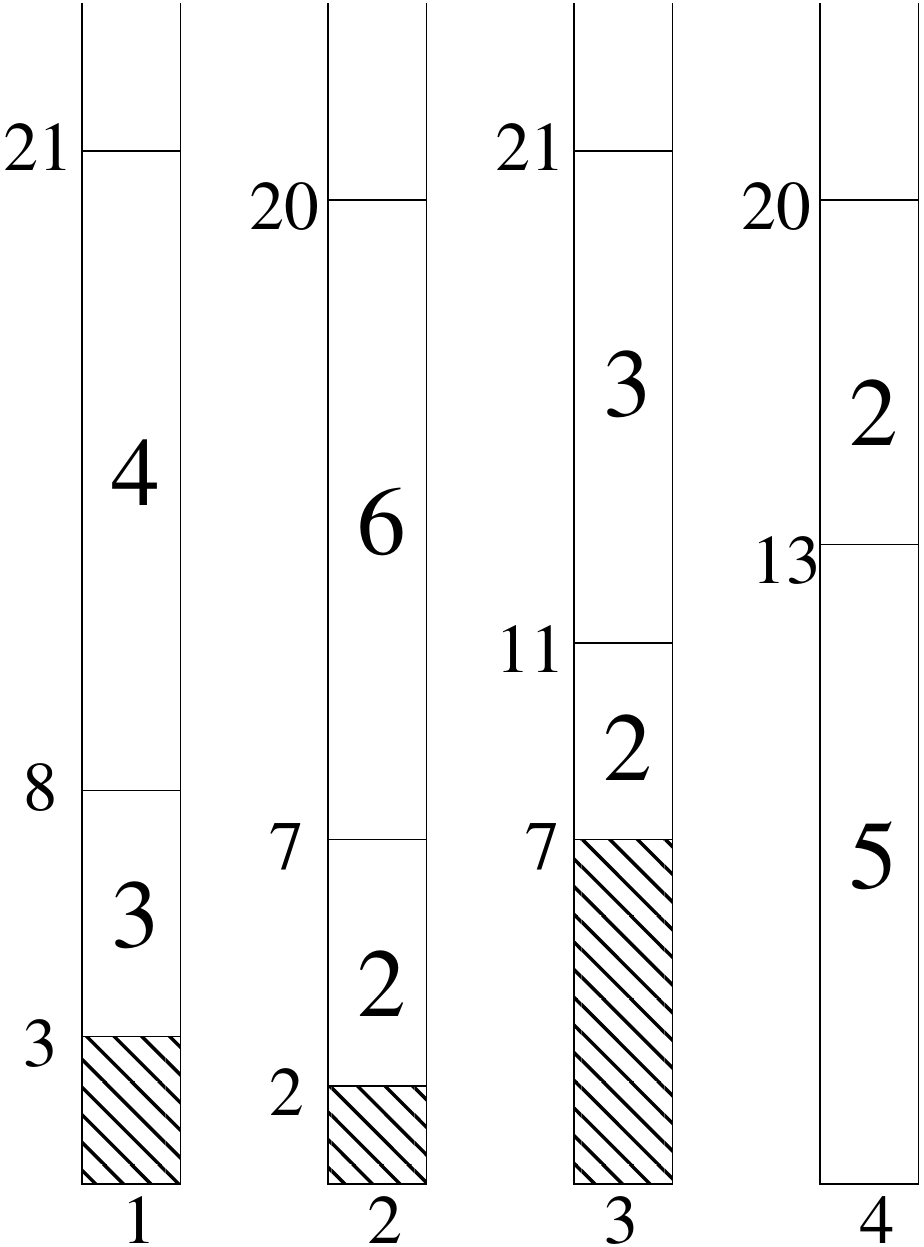}
\caption{A globally optimal schedule that respects a non-optimal distribution 3}
\label{fig2e}
\end{subfigure}
\hspace{0.1 cm}
\begin{subfigure}[b]{0.4\linewidth}
\centering
\includegraphics[width=0.5\linewidth]{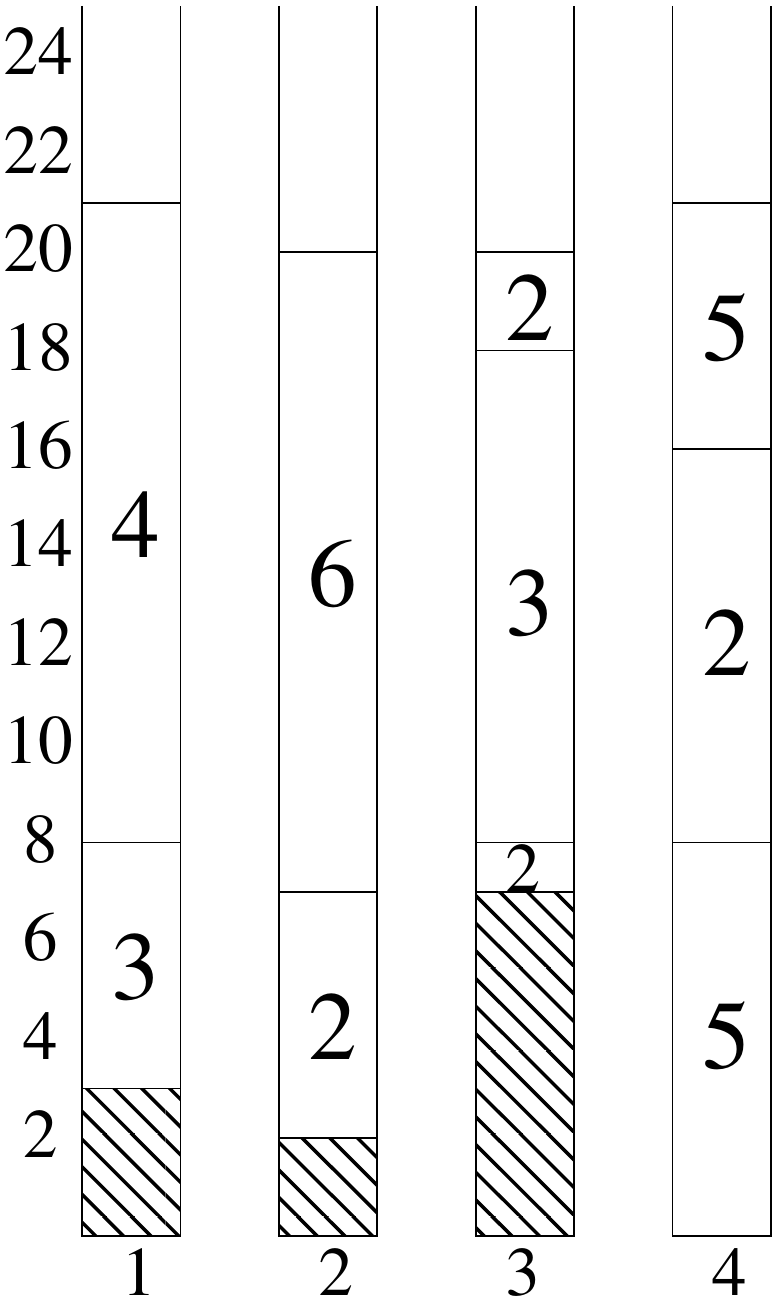}
\caption{A globally optimal schedule that respects a non-optimal distribution 4}
\label{fig2f}
\end{subfigure}
\caption{}
\label{fig2}
\end{figure}

There are a few optimal distributions to linear program LP2$(C_{\max })$ with 
$C_{\max }=18$. We consider two of them which assign (preempted) parts of jobs 3 
and 2 to machine 3. In distribution 1, the processing times  are as follows: 
\\ $t_{13}=5,\ t_{14}=13$, $t_{22}=4, \ t_{26}=13$, 
$t_{32}=8,\ t_{33}=10$ and $t_{45}=13,\ t_{42}=4$.\\ Note that this is not 
an optimal distribution to linear program LP3$(C_{\max })$ due to inequalities 
(\ref{2'}) where $C_{\max }=r_4+p_{14}=8+13=21$ is attained for job $4$. A
non-feasible schedule respecting distribution 1 is depicted in Fig. 2a, and an
optimal schedule respecting the same distribution with makespan 24 is 
depicted in Figure 2b.
  
An optimal distribution 2 is identical to distribution 1 except that 
$t_{22}=5, \ \ t_{32}=6$ and $t_{42}=5$ (see Fig. 2c).  An optimal schedule 
respecting this distribution has the makespan 23, see Fig. 2d.

As can see, distribution 2 possesses better  properties, so that 
an optimal schedule respecting that distribution has a smaller makespan  
than an optimal schedule respecting distribution 1. At the same time, none 
of these schedules is (globally) optimal (see below). Moreover, one 
can easily verify that there exists  no optimal schedule respecting any 
optimal distribution to linear program LP2$(C_{\max })$.  
(From here on we refer to a schedule with minimum makespan 
respecting a given optimal distribution as an {\em  optimal schedule respecting 
that distribution}; such schedule, may not be (globally) optimal, i.e., there
may exist no optimal schedule respecting this distribution. Moreover, 
there may exist no optimal schedule respecting any optimal distribution.) 

Now we consider a  modification of the above considered distributions, a
non-optimal distribution 3 with $C_{\max }=20$, in which $t_{22}=5$, $t_{32}=4$, 
$t_{42}=7$. 
A schedule with the makespan 21 respecting this distribution is depicted in 
Fig. 2e. Distribution 3 is not optimal for linear program 
LP2$(C_{\max })$ and it is not feasible to linear program LP3$(C_{\max })$ 
(e.g., $r_4+t_{14}=21>20$, see inequalities \ref{2'}).  It is 
easy to see that the schedule of Fig. 2e is (globally) optimal. 

In Fig. 2f another optimal schedule with the makespan 21 is depicted. This 
schedule respects distribution 4 with $C_{\max }=21$, which is not optimal for
linear program LP2$(C_{\max })$ but it is (feasible and) optimal for linear 
program LP3$(C_{\max })$ (due to inequalities (\ref{2'}), $C_{\max }=8+13=21$ 
is attained for job $4$, and 21 is also the load of machine 4). However, this 
schedule is not feasible for an instance of 
$R |r_j, pmtn \ - \ no split |L_{\max}$. Distribution 4 
differs from the above optimal distributions on machines 3 and 4. 
Job processing times are distributed as follows:\\ $t_{13}=5,\ t_{14}=13$, 
$t_{22}=5, \ t_{26}=13$, $t_{32}=3,\ t_{33}=10$ and $t_{45}=13,\ t_{42}=8$.\eop

As we saw from the above example, optimal schedules respecting different optimal 
distributions may have different makespan. ``Guessing'' an optimal distribution 
and also a ``suitable'' linear program is an important and also difficult 
task. Furthermore, as we argue in the next section, unlikely, there exists a 
universal ``suitable''
linear program for the studied scheduling problem, such that an optimal 
schedule respecting an optimal distribution to that linear program can be 
guaranteed to be (globally) optimal. Below we give a smaller problem instance
illustrating similar points.

\medskip

\begin{figure}[ht!]
\centering
\begin{subfigure}[b]{0.35\linewidth}
\centering
\includegraphics[width=0.3\linewidth]{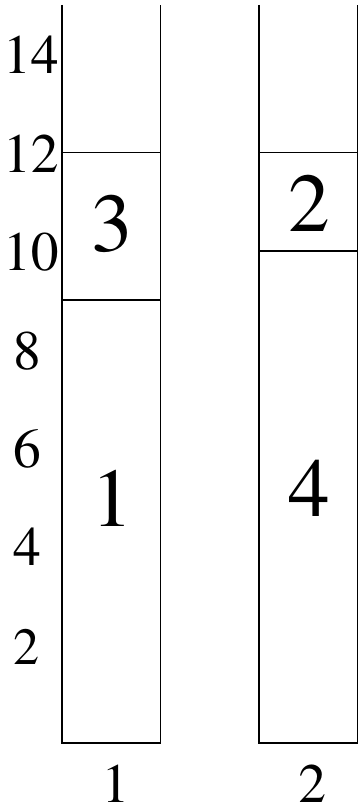}
\caption{} 
\label{fig3a}
\end{subfigure}
\hspace{0.3 cm}
\begin{subfigure}[b]{0.35\linewidth}
\centering
\includegraphics[width=0.3\linewidth]{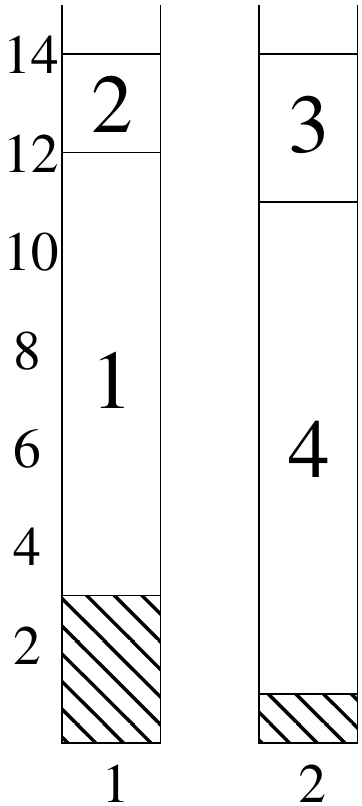}
\caption{}
\label{fig3b}
\end{subfigure}
\caption{A non-feasible schedule respecting an optimal distribution (a) and
an optimal schedule (b)}
\label{fig3}
\end{figure}

\medskip

{\bf Example 2.} As another example, consider a smaller problem instance with four 
jobs on two machines. Jobs 1 and 4 are released at times 3 and 1, respectively, 
and jobs 2 and 3 are released behind these jobs at time, say 5, i.e.,\\
$r_1=3, \ r_2=1$ and $r_3=r_4=5$.\\ Job processing times are as follows:\\
$p_{11}=9,\ p_{12}=\infty$, $p_{14}=\infty,\ p_{24}=10$,
$p_{13}=p_{23}=3$ and $p_{12}=p_{22}=2$. \\ An optimal distribution 
with $C_{\max}=12$ defines the processing times $t_{11}=9, \ t_{13}=3$
and $t_{24}=10,\ t_{22}=2$. This distribution is optimal for both linear 
programs LP2$(C_{\max })$ and LP3$(C_{\max })$. A non-feasible schedule with 
makespan 12 respecting this distribution is depicted in Figure 3a, whereas an optimal
feasible schedule with makespan 14 is depicted in Figure 3b. The latter schedule
respects another distribution with processing times $t_{11}=9, \ t_{12}=2$ and 
$t_{24}=10,\ t_{23}=3$ (in which the roles of jobs 2 and 3 are interchanged) and
it is not optimal for linear programs LP2$(C_{\max })$ and LP3$(C_{\max })$. 
The latter distribution is however optimal for linear program MILP$(C_{\max})$ 
that we introduce in the next section.\eop

\section{Another linear program}

Linear program LP3$(C_{\max })$, although it properly reflects 
the desired restriction for each job, it does not reflect actual 
restrictions imposed by job release times on machine start times  
due to the nature of inequalities (\ref{1}) dealing with a mere sum of 
processing times of jobs assigned to each machine. This causes potential 
conflicts while scheduling machines. In particular, there 
may exist no optimal schedule to problem $R |r_j; pmtn |C_{\max}$ respecting 
an optimal distribution to linear program LP3$(C_{\max })$. 
An appropriate modification of restrictions (\ref{1}) would require a kind of 
``prediction'' of an actual completion time of each machine given the job 
parts assigned to this and other machines possessing parts of the same jobs.
Such a prediction however seems to be unrealistic since it would actually 
require the outcome of scheduling process on each machine. Nevertheless,
we may stall make some assumptions on the scheduling strategy on each machine. 
In particular, we easily observe that no avoidable gap is created on any machine 
in an optimal schedule. It is easy to see that, among the job parts assigned to 
a machine, the first included one corresponds to an earliest released job in an 
optimal schedule. In general, whenever an idle time is unavoidable on a machine, 
an earliest released job is scheduled immediately after that idle time on that 
machine. However, in a liner program with these kind of restrictions the use of
Boolean variables seems to be unavoidable. 

Define $nm$ 0-1 variables, $z_{ij}$ being 1 if a (non-empty) part of 
job $j$ is assigned to machine $i$, and 0 otherwise. Let, further, $J(>r)$ denote
the set of jobs having the release time no smaller than $r$. Then we can rewrite 
$m$ constraints (\ref{1}) into $nm$ constraints as follows.

\begin{equation}\label{1'}
 z_{ij}r_j +  \sum\limits_{l\in J(>r)} x_{il}p_{il}\le C_{\max}, \ \ \
 j=1,\dots,n, \ \ \ i=1,\dots ,m
\end{equation}
 
By incorporating these restrictions instead of constraints (\ref{1}) and 
introducing additional $nm$ 0-1 constraints $$z_{ij}\in\{0,1\}$$ into 
linear program LP3$(C_{\max })$, we obtain a  mixed 0-1 integer linear program 
MILP$(C_{\max})$.

\begin{figure}[ht!]
\centering
\begin{subfigure}[b]{0.4\linewidth}
\centering
\includegraphics[width=0.5\linewidth]{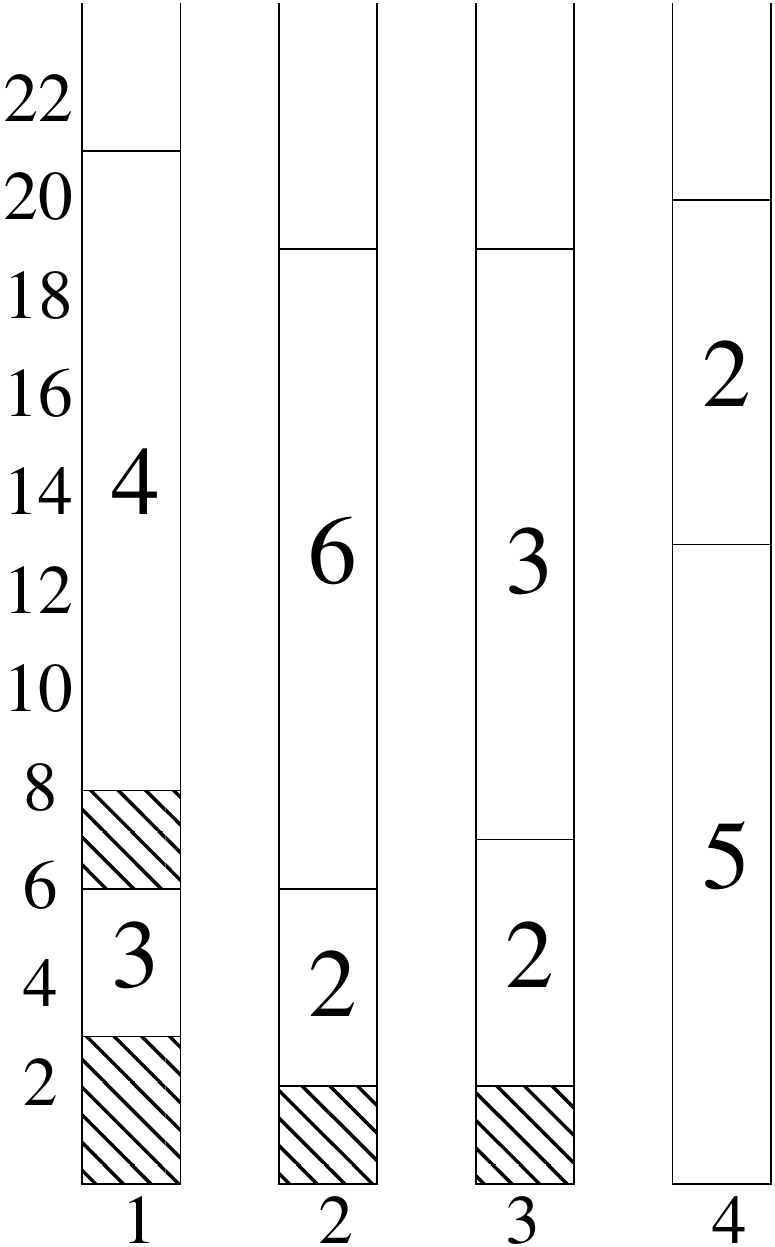}
\caption{A non-feasible schedule respecting the first optimal distribution} 
\label{fig4a}
\end{subfigure}
\hspace{0.1 cm}
\begin{subfigure}[b]{0.4\linewidth}
\centering
\includegraphics[width=0.5\linewidth]{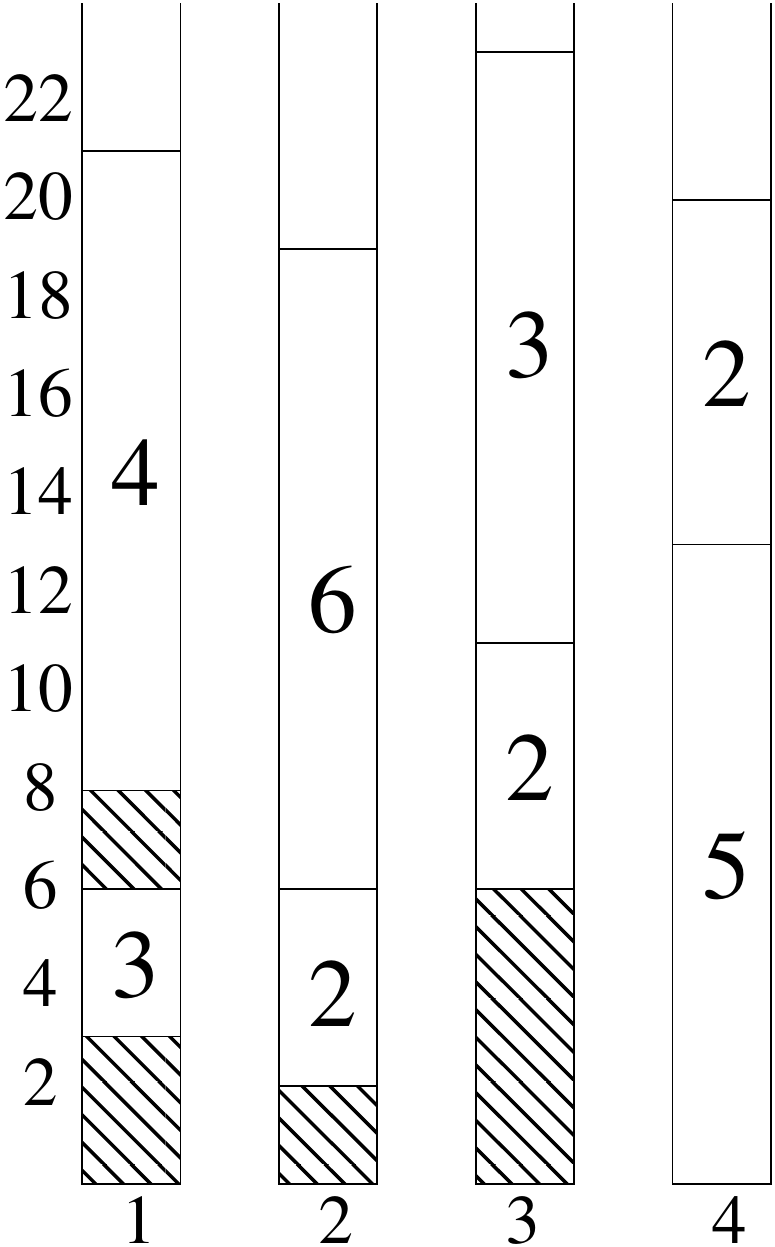}
\caption{A feasible schedule respecting the first distribution}
\label{fig4b}
\end{subfigure}
\hspace{1.2 cm}
\begin{subfigure}[b]{0.25\linewidth}
\centering
\includegraphics[width=\linewidth]{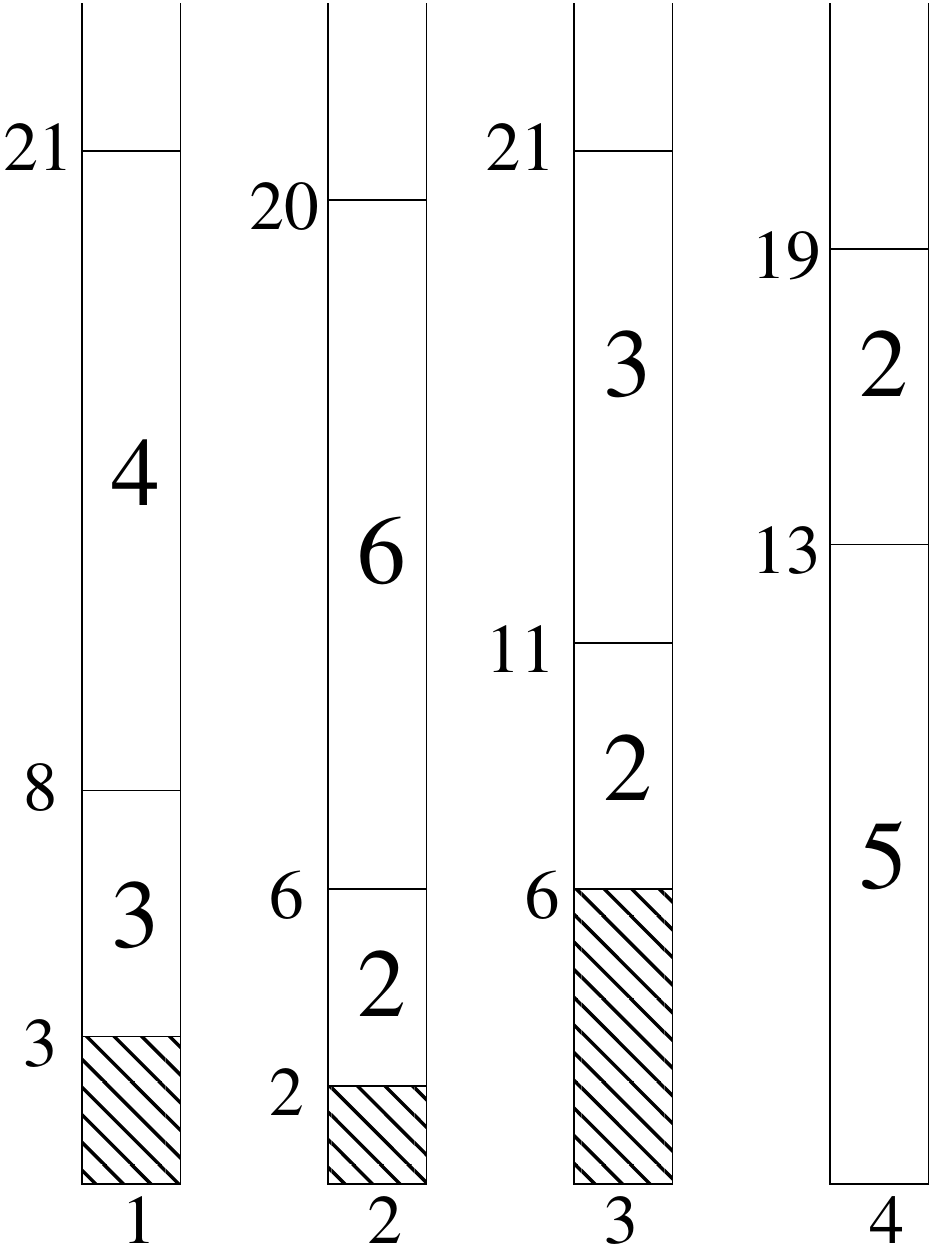}
\caption{A globally optimal schedule respecting the second optimal distribution} 
\label{fig4c}
\end{subfigure}
\hspace{1.2 cm}
\begin{subfigure}[b]{0.25\linewidth}
\centering
\includegraphics[width=\linewidth]{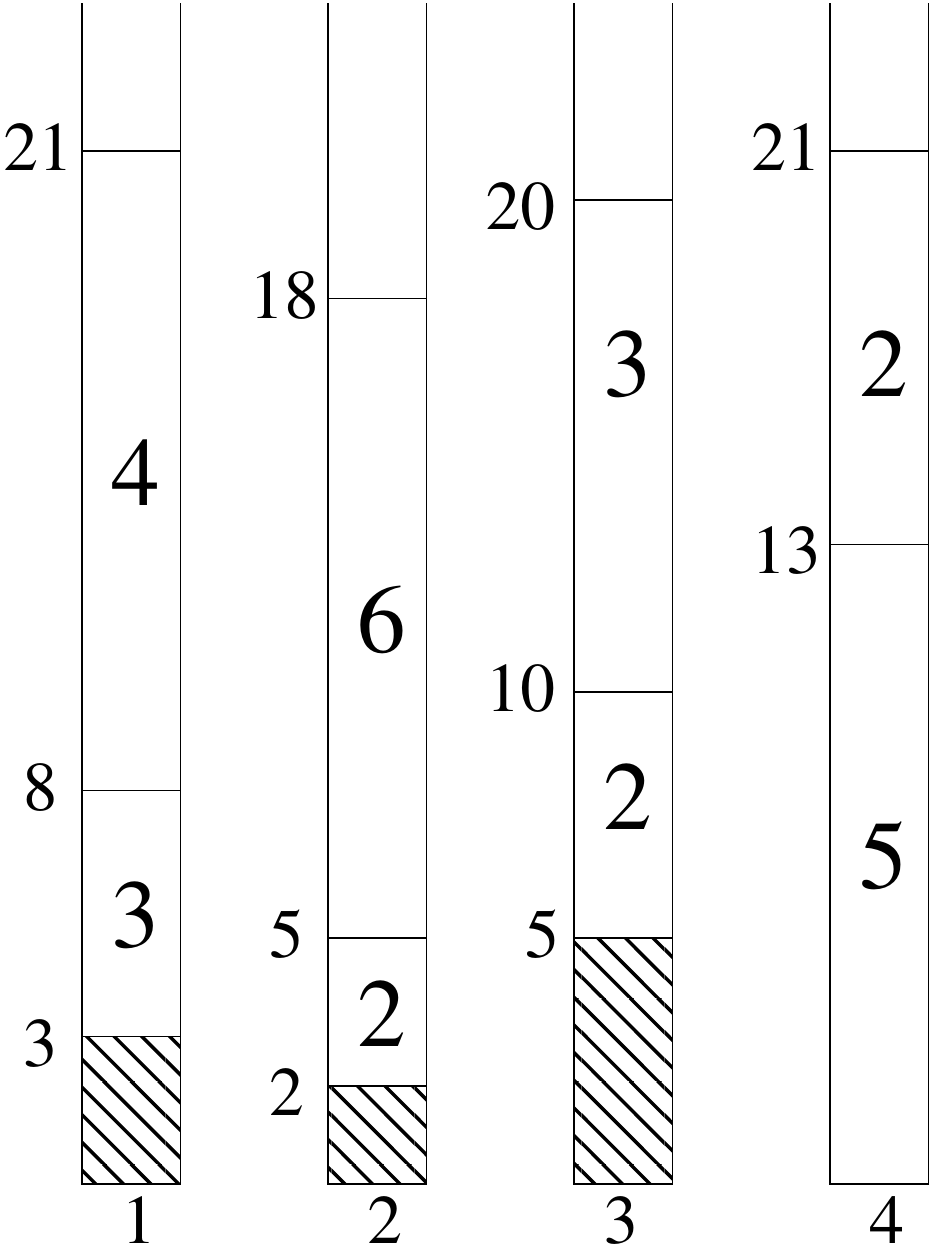}
\caption{A globally optimal schedule respecting the third optimal distribution}
\label{fig4b}
\end{subfigure}
\caption{Schedules constructed based on distributions to linear program MILP$(C_{\max})$.}
\label{fig4}
\end{figure}

\medskip

{\bf Example 1 (continuation).} Returning to the problem instance of Example 1,   
we can easily observe that, for an optimal distribution to the new linear program 
MILP$(C_{\max})$,  $C_{\max}=21$. In particular, distribution 4 from the previous 
section  (Fig. 2f) is optimal also for linear program MILP$(C_{\max})$. 
There exist other optimal distributions 
to linear program MILP$(C_{\max})$. In one of them, \\
$t_{13}=3,\ t_{14}=13$, $t_{22}=4, \ t_{26}=13$, $t_{32}=5,\ t_{33}=12$ and 
$t_{45}=13,\ t_{42}=7$, \\ see Fig. 4a representing a non-feasible schedule 
that respects this distribution. A feasible  schedule with makespan 23 
respecting the same distribution is depicted in Fig. 4b. 
In a slight modification of the latter optimal distribution, job 3 is redistributed 
on machines 1 and 3 so that its processing time on machine 1 is increased by 
2 and its processing time on machine 3 is decreased by the same amount. This
results in a globally optimal schedule with makespan 21 depicted in Fig. 4c. 
In another optimal distribution $t_{22}$ is reduced to $3$ and $t_{42}$ 
is increased to $8$. Another globally optimal schedule with makespan 21 respecting 
the latter optimal distribution is depicted in Fig 4d.\eop  

As we can see,  even for a very small sized problem instance, a number of
different optimal distributions to the new linear program MILP$(C_{\max})$
exist, some of them leading to an optimal schedule and some not.  We again
need to ``guess'' a ``correct'' optimal distribution among all possible optimal 
ones. Moreover, as we show below, not necessarily there exists a globally optimal 
schedule that respects an optimal distribution to the new linear program 
MILP$(C_{\max})$,  as it was the case for liner programs 
LP2$(C_{\max })$ and LP3$(C_{\max })$.

\medskip

{\bf Example 1a.} Consider a slight modification of the problem instance of 
Example 1 in which all job parameters remain the same except that $r_4=0$. This 
reduces the makespan of an optimal distribution to linear program MILP$(C_{\max})$
from $C_{\max}=21$ to  $C_{\max}=20$. Now job 4 can be partitioned on
machine 1 in two parts, hence $t_{13}$ can be increased to 7. An optimal schedule 
with makespan 20 respecting this optimal distribution is depicted in Fig. 5a.\eop 

{\bf Example 1b.} For the second modification of Example 1, let $r_4=6$.  
For an optimal distribution to this modified instance $C_{\max}=20$, which 
is the completion time of machine 4 (note that the completion time on machine 1, 
compared to that in the schedule of Fig. 4b, is reduced by the length of the gap 
$[6,8)$). A non-feasible schedule respecting an optimal distribution  is 
depicted in Fig. 5b. An optimal feasible schedule with makespan 23 respecting 
the same distribution is depicted in Fig. 5c. The latter schedule is not globally 
optimal. An optimal schedule with makespan 21 is illustrated in Fig 5d; this 
schedule respects a distribution with the same makespan $C_{\max}=21$. Observe that
the latter distribution is not optimal for linear program MILP$(C_{\max})$.\eop

\begin{figure}[ht!]
\centering
\begin{subfigure}[b]{0.2\linewidth}
\centering
\includegraphics[width=\linewidth]{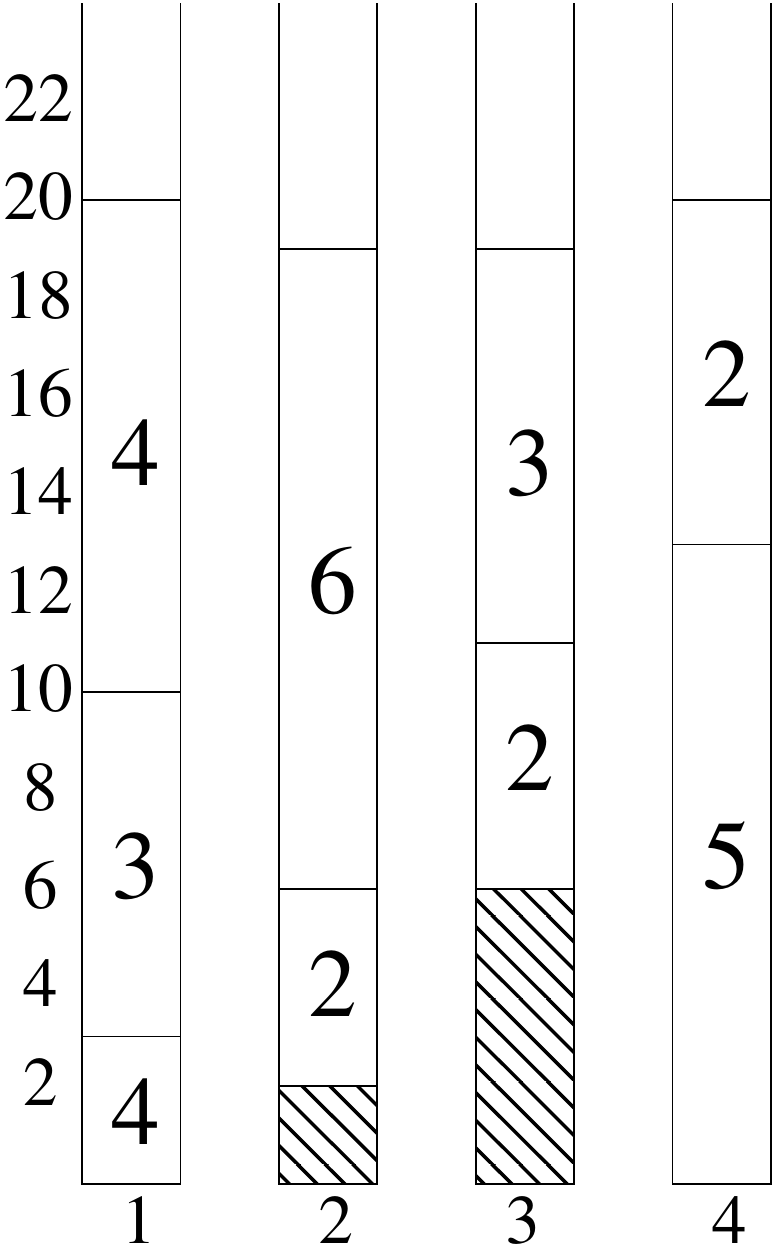}
\caption{} 
\label{fig45a}
\end{subfigure}
\hspace{0.3 cm}
\begin{subfigure}[b]{0.2\linewidth}
\centering
\includegraphics[width=\linewidth]{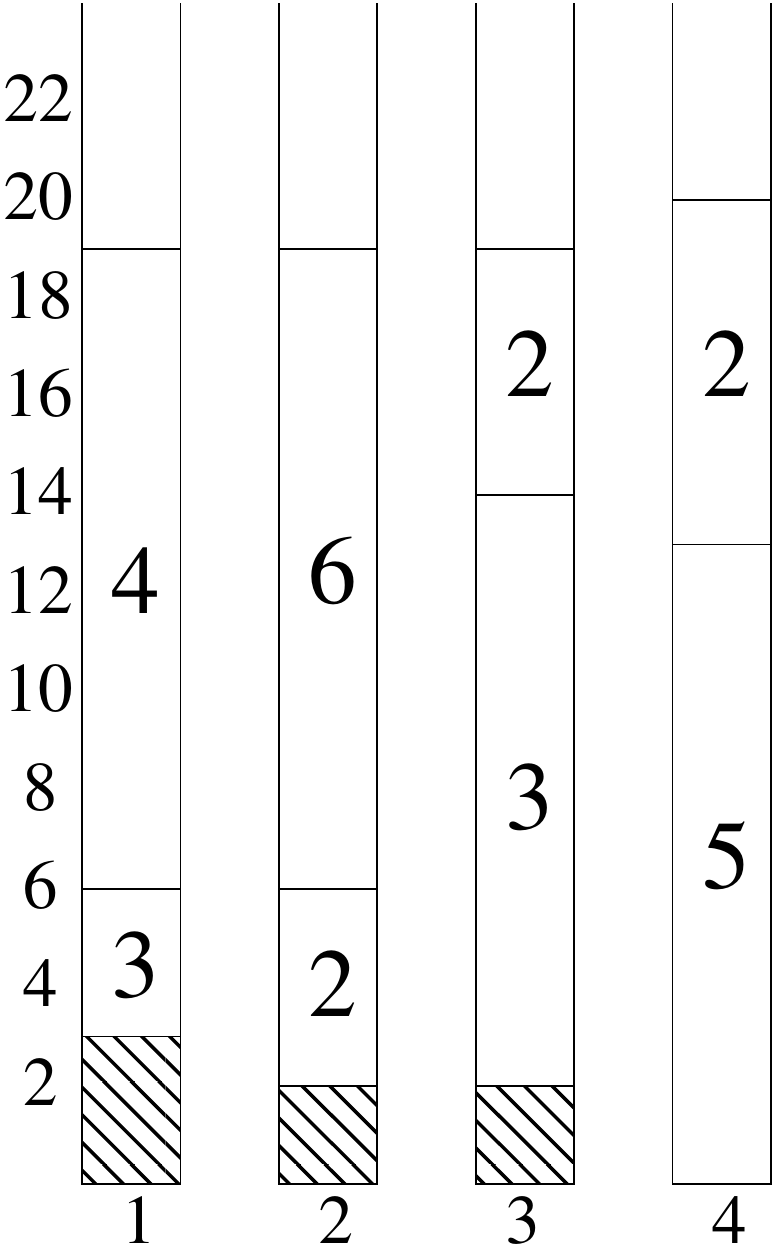}
\caption{}
\label{fig5b}
\end{subfigure}
\hspace{0.3 cm}
\begin{subfigure}[b]{0.2\linewidth}
\centering
\includegraphics[width=\linewidth]{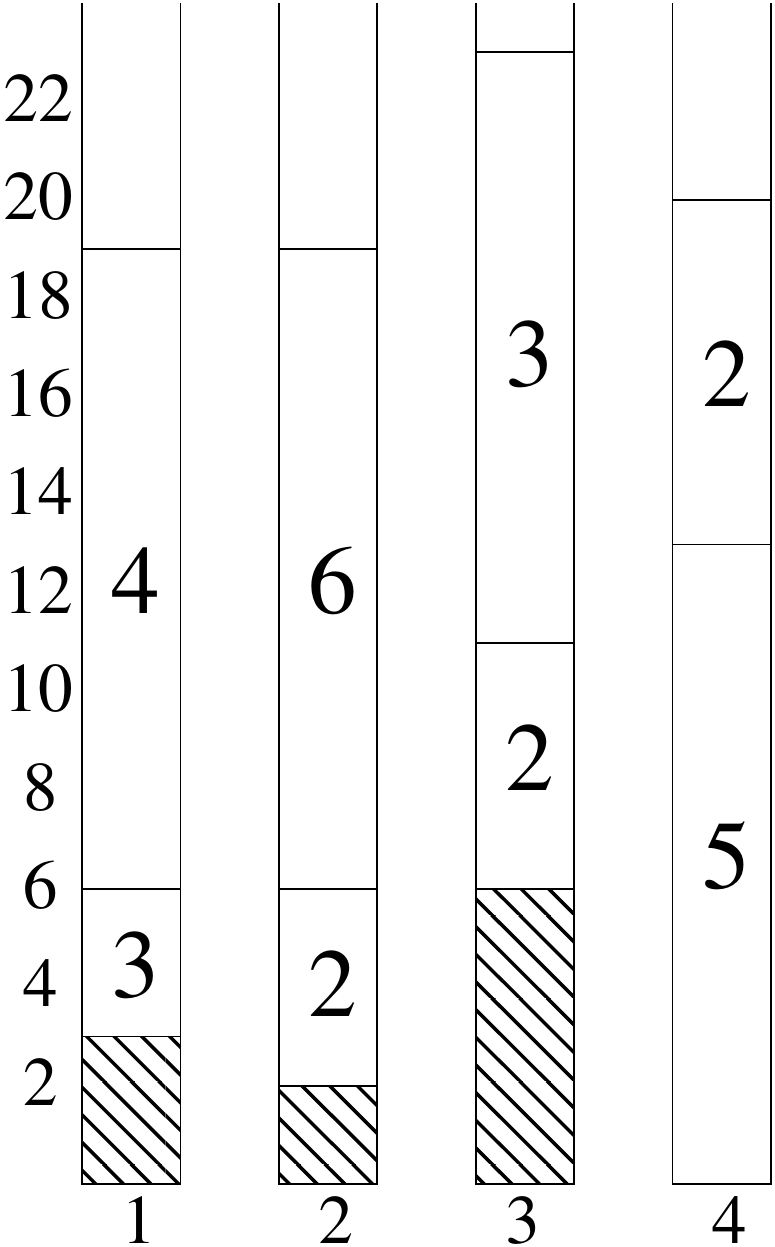}
\caption{} 
\label{fig45c}
\end{subfigure}
\hspace{0.3 cm}
\begin{subfigure}[b]{0.2\linewidth}
\centering
\includegraphics[width=\linewidth]{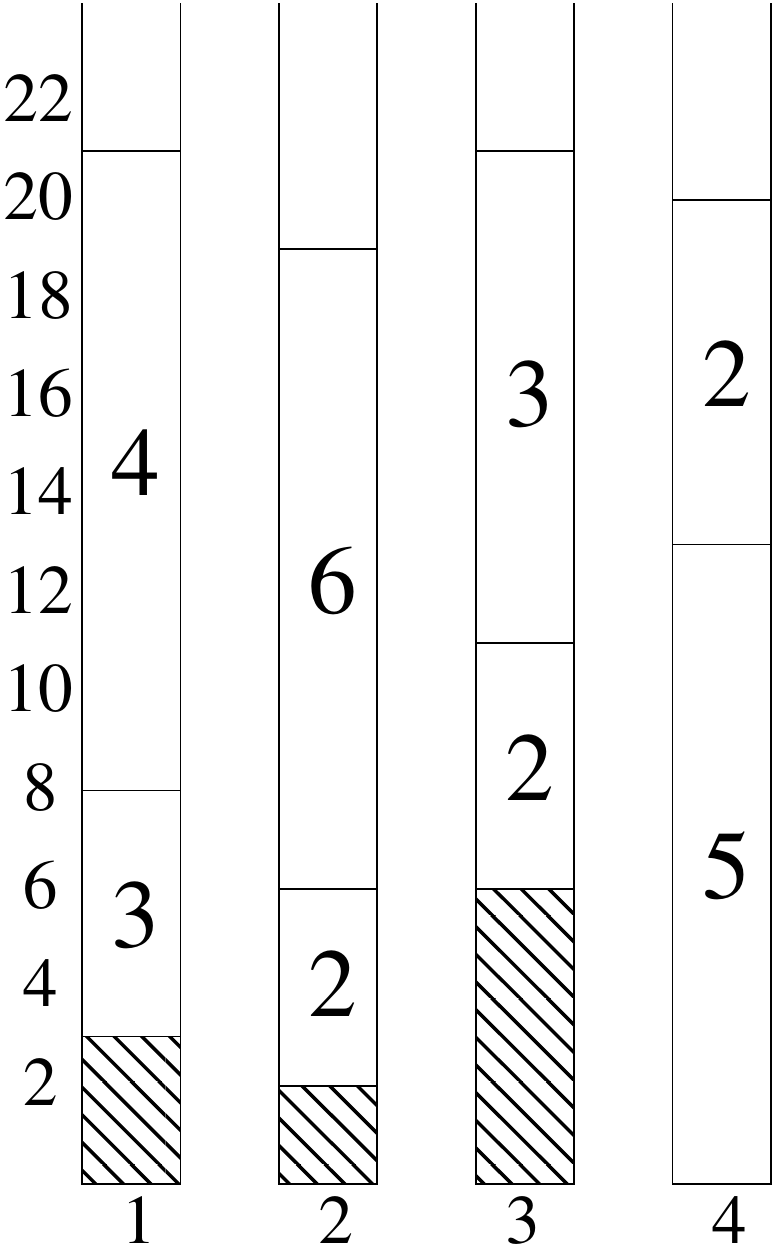}
\caption{} 
\label{fig5d}
\end{subfigure}
\caption{(a): An optimal schedule for the first modified instance; (b): a non-feasible
schedule respecting an optimal distribution for the second modified instance; (c): 
an optimal schedule respecting the latter distribution; (d): and an optimal
schedule, that respects a non-optimal distribution}
\label{fig5}
\end{figure}

\section{The scheduling method of Lawler and Labetoulle}

In the following section we describe the method of Lawler and Labetoulle \cite{ll78} 
for the construction of an optimal schedule from an optimal distribution to linear 
program LP2$(C_{\max })$ for problem $R |pmtn |C_{\max}$, and then we show how it
can be extended to construct an optimal schedule respecting any
distribution to problem $R |r_j; pmtn |C_{\max}$. 

Lawler and Labetoulle \cite{ll78} adopted open shop scheduling technique of 
Gonzalez and Sahni \cite{GSopen} for the construction of an optimal feasible
schedule with makespan $C_{\max }$ respecting an optimal distribution to program 
LP2$(C_{\max })$ for simultaneously released jobs (note that an open shop instance 
can already be seen as a distribution). Recall that an optimal distribution $\{x_{ij}\}$
defines an $m$ x $n$ non-negative processing time matrix $T=\{t_{ij}\}=x_{ij}p_{ij}$. 
The so-called {\em decrementing sets} are iteratively formed from iteration 1 by 
selecting one entry in each tight row and in each tight column: a row/column is 
{\em tight} iff the sum of the entries in that row/column is exactly $C_{\max }$. 
The initial processing time matrix $T=T^1$, defined by an optimal distribution, 
is iteratively transformed into the $m$ x $n$ 0-matrix. At each iteration $h>0$, 
the matrix $T^{h-1}$ of the previous iteration is updated according to the 
formed decrementing set $D^h$ at iteration $h$; in particular, each entry in 
matrix $T^{h-1}$ corresponding to an element of set $D^h$ is decreased by a suitably 
chosen (small enough) number $\delta^h$ resulting in the updated matrix $T^h$ of 
iteration $h$. $\delta^h$ is chosen in such a way that the new matrix $T^h$ possesses
similar properties as its predecessor matrix $T^{h-1}$: The sum 
of the elements in each row and column of the updated matrix is no larger than 
$C_{\max }-\sum_{i=1}^h \delta^i$. The elements in decrementing set $D^h$ define
a partial schedule of iteration $h$ of length $\delta^h$ according to the
selected portions of processing times. $\sigma^h$ is the partial
schedule generated by iteration $h$, which is obtained by a mere merging of the 
partial schedules of each of the iterations $1,\dots,h$, so that the 
makespan of partial schedule $\sigma^h$ is $\tau^{h+1}=\sum_{i=1}^h \delta^i$. 
We will refer to $\tau^{h+1}$ as the {\em scheduling time}  of iteration 
$h+1$, which is the time moment at which the partial schedule of iteration $h+1$ 
starts. The whole procedure halts at iteration $h$ such that matrix $T^h$ is a 
0-matrix. At that iteration, $\sigma^h$ is a complete feasible schedule. The
optimality of this complete schedule immediately follows from the fact that
its makespan is $C_{\max}$ (a lower bound on the optimum schedule makespan). 

In the initial matrix $T$ and in each 
following matrix $T^h$ there exists a decrementing set. This follows from 
a known Birkhoff and von-Neumann theorem stating that every doubly stochastic 
matrix is a linear combination of permutation matrices. 
Indeed, by completing matrix $T$ with additional slack rows and columns, 
an $m+n$ x $m+n$ matrix with the entries of {\em each} its row and column summing up 
to $(C_{\max })$ can be obtained. Dividing all the entries of this matrix by 
$(C_{\max })$, a doubly stochastic matrix is obtained. Then a permutation matrix 
from the theorem defines a decrementing set. 

\medskip

\begin{figure}[ht!]
\centering
\begin{subfigure}[b]{0.2\linewidth}
\centering
\includegraphics[width=\linewidth]{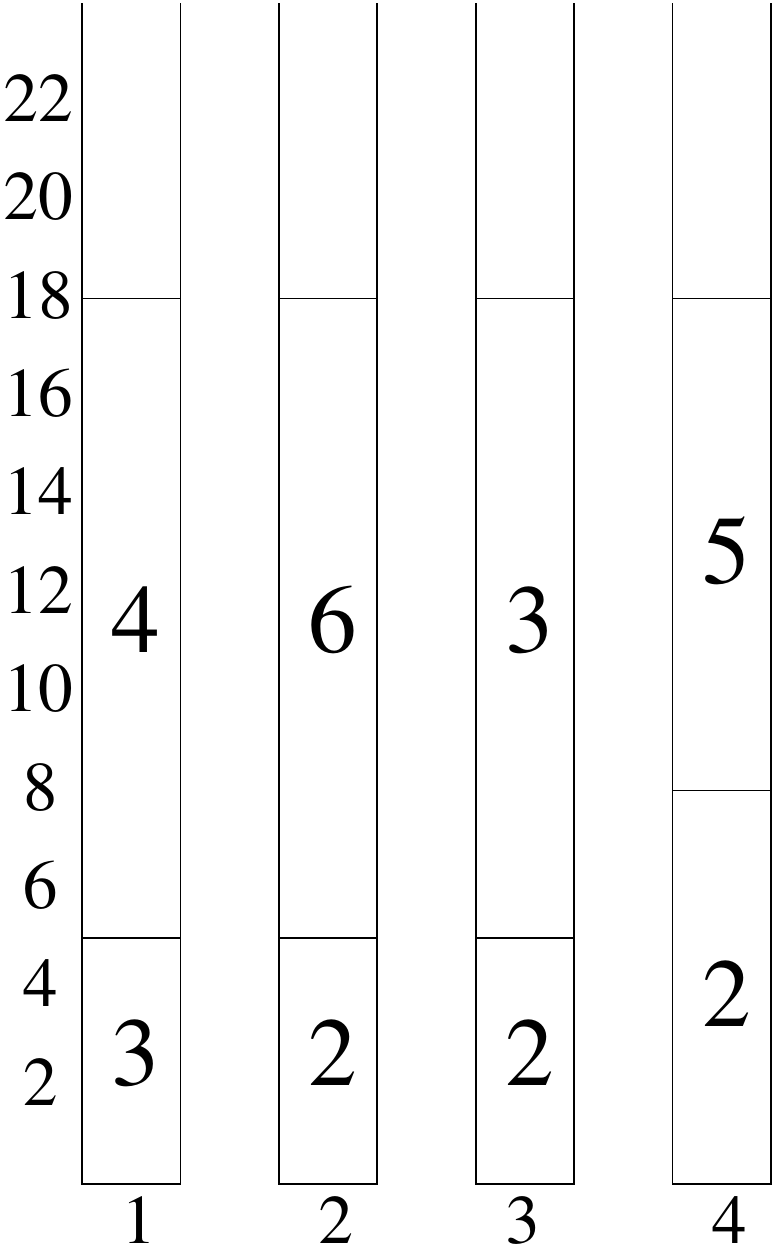}
\caption{} 
\label{fig5a}
\end{subfigure}
\hspace{0.6 cm}
\begin{subfigure}[b]{0.2\linewidth}
\centering
\includegraphics[width=\linewidth]{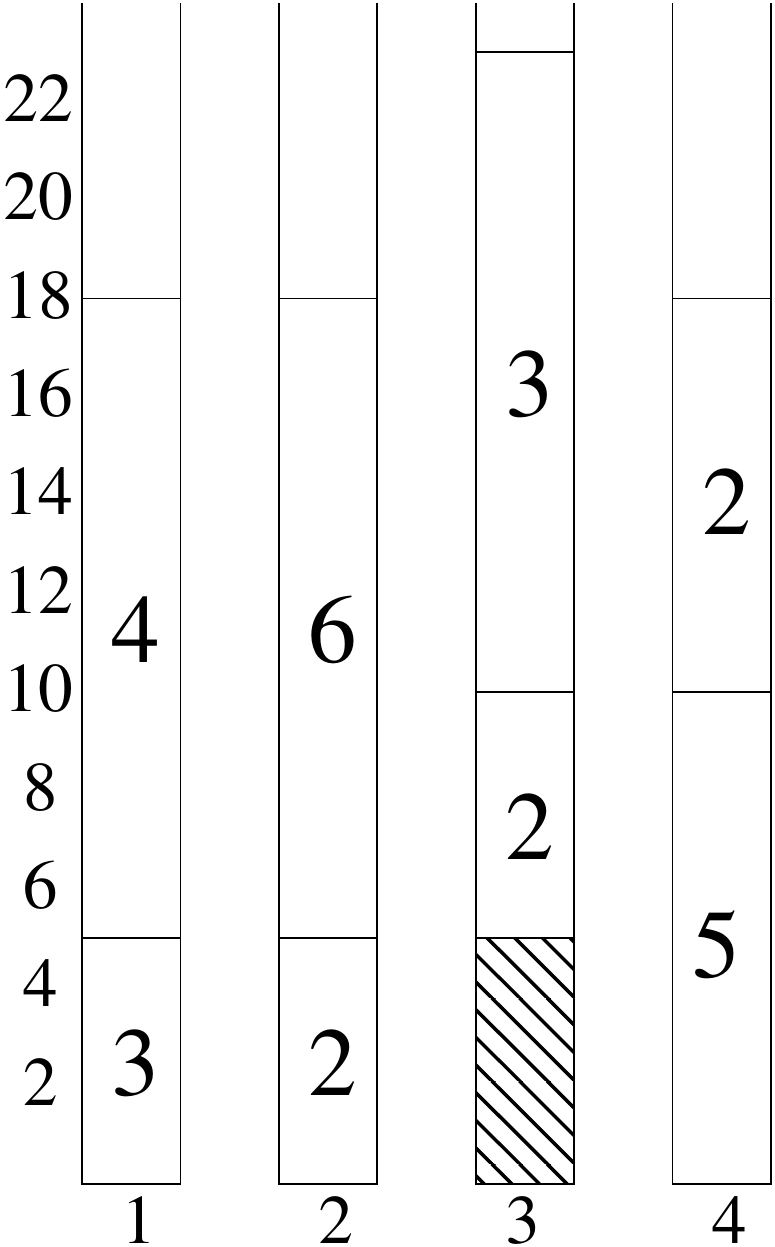}
\caption{}
\label{fig5b}
\end{subfigure}
\hspace{0.6 cm}
\begin{subfigure}[b]{0.2\linewidth}
\centering
\includegraphics[width=\linewidth]{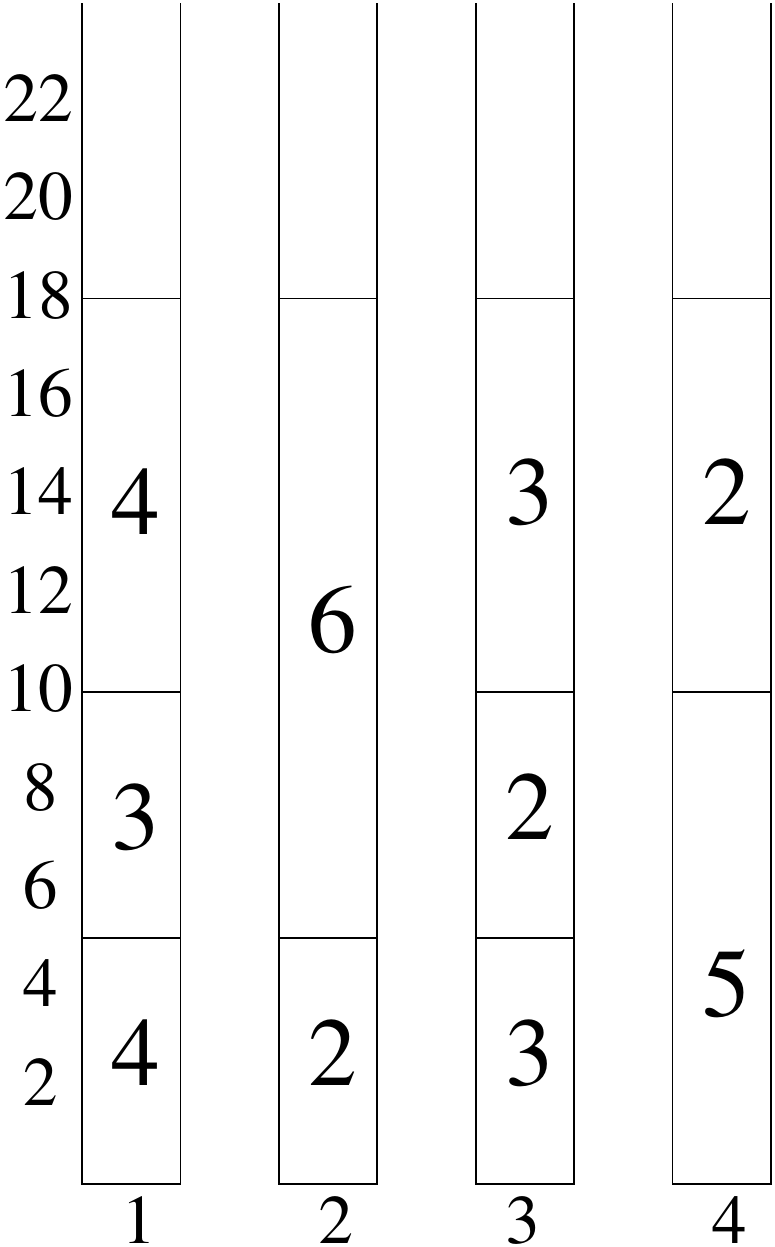}
\caption{} 
\label{fig5c}
\end{subfigure}
\hspace{0.6 cm}
\begin{subfigure}[b]{0.2\linewidth}
\centering
\includegraphics[width=\linewidth]{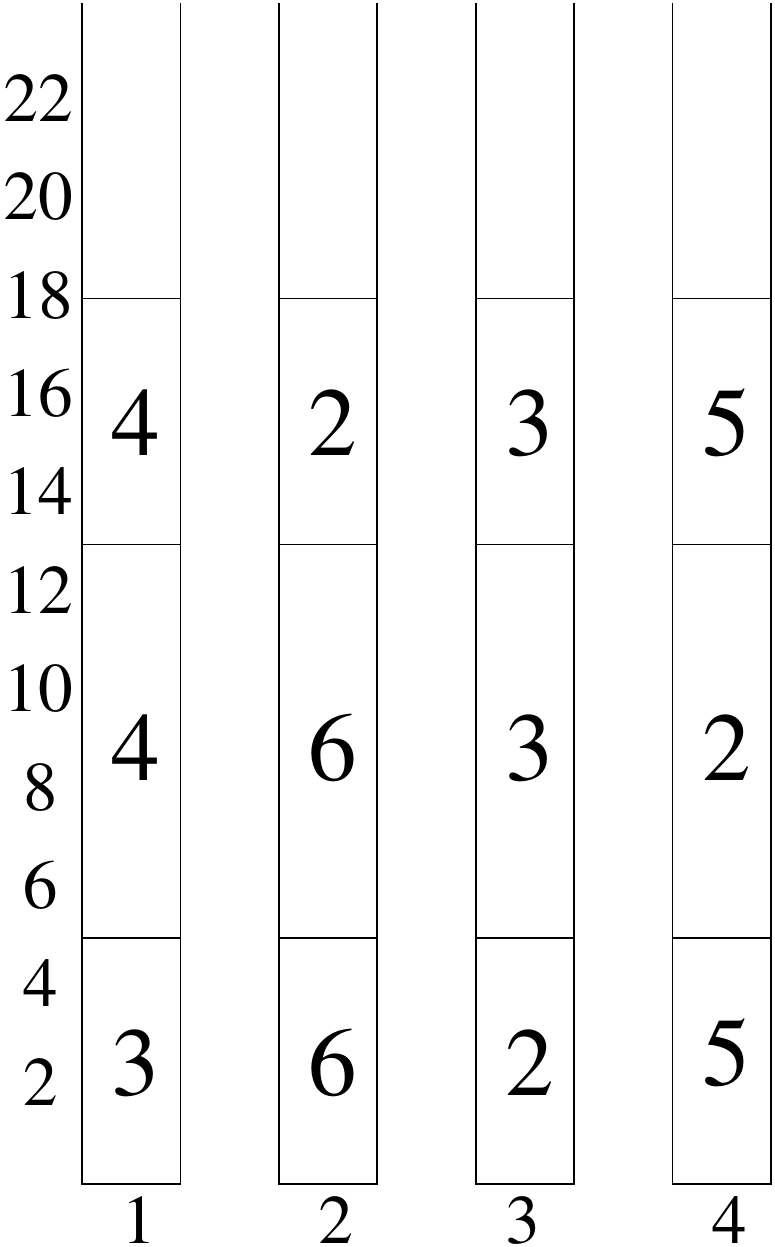}
\caption{}
\label{fig5d}
\end{subfigure}
\caption{A non-feasible schedule respecting an optimal distribution (a), a feasible 
non-optimal schedule (b) and two optimal schedules (c) and (d) respecting the 
same distribution.}
\label{fig6}
\end{figure}

{\bf Example 4.} Let us illustrate the scheduling method of Lawler and Labetoulle
on a small instance of problem $R |pmtn |C_{\max}$. It is basically the instance of 
Example 1 adopted for the case when all jobs are simultaneously released. To maintain
$C_{\max}=18$, we increase the processing time of jobs 2 and 3 to 18 and decrease 
the processing time of job 5 to 10. It can be easily verified that in an optimal 
distribution to linear program LP2$(C_{\max})$ with $C_{\max}=18$, we have
$t_{13}=5,\ t_{14}=13$, $t_{22}=5, \ t_{26}=13$, $t_{32}=5, \ t_{33}=13$, and 
$t_{42}=8, \ t_{45}=10$. A non-feasible schedule respecting this distribution is
depicted in Fig. 6a. This schedule can straightforwardly be converted to a feasible
schedule of Fig. 6b with makespan 23 by imposing a gap $[0,5)$ on machine 3. This
schedule is not optimal. An optimal one respecting the above
optimal distribution is depicted in Fig. 6c.

\begin{table}
\centering
\caption{Flowchart of the procedure with processing time matrices and 
the corresponding decrementing sets (marked with circles)} \label{table1}
\resizebox{17cm}{!}{%
\begin{tabular}{cccccc}

\begin{tabular}{c}
  \\ 
1 \\ 
2 \\ 
3 \\ 
4 \\ 
\end{tabular}
& 
\begin{tabular}{|c|c|c|c|c|}
\hline 
2 & 3 & 4 & 5 & 6 \\ 
\hline 
0 & \textcircled{5} & 13 & 0 & 0 \\ 
\hline 
5 & 0 & 0 & 0 & \textcircled{13} \\ 
\hline 
\textcircled{5} & 13 & 0 & 0 & 0 \\ 
\hline 
8 & 0 & 0 & \textcircled{10} & 0 \\ 
\hline 
\end{tabular}
&
\begin{tabular}{|c|c|c|c|c|}
\hline
2 & 3 & 4 & 5 & 6 \\ 
\hline 
0 & 0 & \textcircled{13} & 0 & 0\\ 
\hline 
5 & 0 & 0 & 0 & \textcircled{8}\\ 
\hline 
0 & \textcircled{13} & 0 & 0 & 0\\ 
\hline 
\textcircled{8} & 0 & 0 & 5 & 0 \\ 
\hline 
\end{tabular}
&
\begin{tabular}{|c|c|c|c|c|}
\hline
2 & 3 & 4 & 5 & 6 \\ 
\hline 
0 & 0 & \textcircled{5} & 0 & 0 \\ 
\hline 
\textcircled{5} & 0 & 0 & 0 & 0\\ 
\hline 
0 & \textcircled{5} & 0 & 0 & 0\\ 
\hline 
0 & 0 & 0 & \textcircled{5} & 0 \\ 
\hline 
\end{tabular}
& 
\begin{tabular}{|c|c|c|c|c|}
\hline
2 & 3 & 4 & 5 & 6 \\ 
\hline 
0 & 0 & 0 & 0 & 0 \\ 
\hline 
0 & 0 & 0 & 0 & 0 \\ 
\hline 
0 & 0 & 0 & 0 & 0 \\ 
\hline 
0 & 0 & 0 & 0 & 0 \\ 
\hline 
\end{tabular}
\\
\end{tabular} 
}
\end{table}

The scheduling method of Lawler and Labetoulle \cite{ll78} will generate an optimal
schedule  as follows. The initial matrix $T=T^1$ of job processing times defined
by the optimal distribution is represented in the first quarter of Table 1, where
the entries corresponding to the elements of the decrementing set $D^1$ and
the following decrementing sets are circled. 
$\delta^1$ can be chosen to be equal to 5. The updated matrix $T^2$ and the entries 
corresponding to the elements of the decrementing set $D^2$ are shown in the second 
quarter of Table 1. The first partial schedule with length 5 can be seen as the initial 
(first) part of the schedule of Fig. 6d corresponding to interval $[0,5)$. 
The computations in the following iterations $h=2,3$ with $\delta^2=8$, $\delta^3=5$
are reflected in the next quarters of Table 1 and in the upper parts in 
the schedule of Fig. 6d. Although schedule $\delta^3$ of Fig. 6d  with makespan
$C_{\max}=5+8+5=18$ is somewhat similar to that of Fig. 6c,  it splits job
5 on machine 4 (hence it is not a feasible solution to problem 
$R |r_j, pmtn \ - \ no split |L_{\max}$).\eop

\subsection{Scheduling non-simultaneously released jobs}

One naturally wishes to extend the schedule construction technique of 
Lawler and Labetoulle \cite{ll78} for non-simultaneously released jobs. 
As we saw, an optimal schedule respecting an optimal distribution 
to any of linear programs that we considered is not necessarily optimal. 
Without giving a formal proof that there may exist no other ``proper'' 
linear program for the problem, we argued in Section 6.1 that
it is unlikely that such linear program exists. As we also observed, 
there may exist no globally optimal schedule respecting an optimal
distribution, i.e., any such schedule may respect a non-optimal distribution. 

Due to these observations, now we would like to find a schedule with the 
minimum makespan among all schedules respecting a given (not necessarily
optimal) distribution, i.e., find an optimal schedule respecting that 
distribution. The schedule construction technique of Lawler and Labetoulle 
\cite{ll78} can be generalized by maintaining an extra information on which 
job parts can be scheduled at every scheduling time $\tau^h$. Note that 
such care is to be taken only on the first scheduled part of each job. For 
that, we introduce an additional row 0 in processing time matrices. 

Initially, in the extended matrix $T^1$, the entry $t^1_{0j}$ in column $j$ 
of row 0 is $r_j$; iteratively, $t^h_{0j}:=\max\{0, r_j-\tau^{h}\}$.
During the scheduling process, we impose an additional restriction that 
forbids scheduling of job $j$ at time $\tau^h$ if $t^h_{0j}>0$ (independently
of whether the corresponding  entries are from a tight row or tight column).

As a result, not necessarily  a decrementing set will contain one entry from 
a tight row or a tight column. In particular, set $D^h$ will contain no entry 
from a (tight) row $i$ if among yet unscheduled job parts assigned to machine 
$i$ no job is yet released by time $\tau^{h+1}$ (i.e., for any positive entry
in row $i$, the corresponding entry in row 0 is positive); likewise, 
the decrementing set of iteration $h$ will contain no entry from a (tight) 
column $j$ if job $j$ is not yet released by time $\tau^{h+1}$. 

We will refer to a row $i$ from matrix $T^h$ as {\em ready} at iteration $h$ if 
it contains a positive entry $t_{ij}^{h}>0$ such that $t^h_{0j}=0$; we will refer 
to such an entry as {\em valid} for row $i$ at iteration $h$.

Let $t^h_{0j}=0$, and let $M_h(j)$ be the set of machines such that the entry
$t_{ij}^{h}$ is valid for $i\in M_h(j)$. Then the ready rows from set $M_h(j)$
are said to be  {\em conflicting} by job $j$ at iteration $h$. Note that two
or more rows may be conflicting by two or more different jobs. 

A decrementing set $D^h$ of iteration $h$ contains one entry from a ready 
row such that no two entries from the same column are included into that set
(since the same job cannot be scheduled on different machines at a time); 
whenever a ready row $i$ contains an element from a tight column $j$, the 
corresponding part of job $j$ can be selected if entry $t_{ij}$ is valid at 
iteration $h$. 

It might not be possible to select an entry from every ready row at a given 
iteration $h$ since two or more rows may be conflicting by the same jobs in
such a way that all entries cannot be selected. For example, if we have two 
ready rows with valid entries only in, say, column $j$, then only one of these 
entries can be selected; likewise, if there are three ready rows with valid 
entries only in, say, columns $j$ and $j'$, 
then only two of these entries, one from column $j$ and the other from column 
$j'$, can be selected in decrementing set $D^h$. We break ties by selecting 
the corresponding entry from  row $i$ with the maximum current load, i.e., 
with the maximum $\sum_{l=1}^n t_{il}^h$. If such a row contains several
valid entries, then the entry from a column $j$ with the maximum remaining 
processing time, i.e., with the maximum $\sum_{l=1}^m t_{lj}^h$ is selected.
Further ties are broken by selecting an entry from column $l$ such that 
$t^{h-1}_{il}\in D^{h-1}$ (i.e., part of job $l$ was included in the 
decrementing set of the previous iteration). The latter tie
breaking rule avoids unnecessary preemption of an already running job. 
(At every iteration $h$, the rows can be considered in non-increasing order 
of their loads and the corresponding valid entries can be selected from a
column $j$ with the maximum remaining processing time.)

Let $r$ be the minimum positive element in row $0$ at iteration $h$ in 
matrix $T^h$, i.e., $r=\min_j t^h_{0j}$. The jump $\delta^h$ at iteration 
$h$ is now defined as the minimum between $\delta^h$ as defined in Lawler and 
Labetoulle \cite{ll78} and $r$. 

Since there may exist no more than $n$ district job release times, the extended 
method yields an additional term $n$ bounding the number of iterations
and hence has the same polynomial time complexity as the method of Lawler and 
Labetoulle \cite{ll78}. It is not difficult to see that $\tau^{h^*}$ is an optimal 
schedule makespan, where $h^*$ is the last iteration in the procedure. First
note that the extended procedure will work as the basic one from the earliest 
scheduling time $\tau^{h'}$ such that $t_{0j}^{h'}=0$, for all $j=1,\dots,n$, since
the corresponding decrementing sets will contains $m$ elements with exactly one
element from each tight column (by the construction, all entries in column $j$ 
will be valid at any iteration $h$ with $\tau^h \ge r_j$). In general, 
the extended procedure will work as the basic one if at every iteration there 
are $m$ non-conflicting ready rows with an entry in a tight column  so that the 
entries from each tight column are included into the corresponding decrementing 
sets. Otherwise, in case there is an entry in a tight 
column that was not selected at an iteration $h$, the corresponding entry in row 
0 should have been positive, i.e., the corresponding job is not released by the 
current scheduling time $\tau^{h-1}$. In other words, if an entry from a tight 
column $j$ was not included in decrementing set $D^h$ then there is no valid entry 
from that column at iteration $h$, equivalently, $r_j > \tau^{h-1}$. No feasible 
schedule may include such a job at time $\tau^{h-1}$. Likewise, if there are only 
$l<m$ non-conflicting ready rows at iteration $h$, then there are only $l$ jobs 
that can feasibly be scheduled at time $\tau^{h-1}$. Our tie breaking rule will
include an entry corresponding to each of these $l$ jobs in decrementing set $D^h$,  
in total $l$ entries from $l$ rows corresponding to $l$ most loaded machines will
be included. Finally, note that among conflicting rows, ties can easily be broken 
by selecting valid entries from the rows corresponding to most loaded machines.

\medskip

{\bf Example 3.} Let us first illustrate the extended scheduling method for the 
scheduling instance from Theorem \ref{1}. Recall an optimal distribution respected 
by the schedule of Fig. 1. Let us first assume that we have a solution to the 
corresponding partition instance, i.e., we have sets $P_1$ and $P_2$. Then, for 
the sake of simplicity, we can represent all partition jobs in one column marked 
as $P$, see Table 2 below. We update the entries in this column according to the 
made selections. All partition jobs from sets $P_1$ and $P_2$ are scheduled 
consequently without creating any machine idle time in two different 
(aggregated) iterations 2 and 4. In the first quarter of Table 2 an 
initial extended processing time matrix $T^0$ is presented. For this instance, 
all three rows are ready from the beginning of iteration 1. The valid entries
corresponding to the elements in the decrementing set $D_1$ are circled (note 
that the corresponding entries in row 0 are 0). Row 1 and column 1 are tight,  
hence element $t_{11}$ is circled. From rows 2 and 3 elements $t_{24}$ and $t_{32}$
are similarly selected. We have $\delta^1=3$ and $\tau^1=3$. The part of the schedule
of Fig. 1 corresponding to the interval $[0,3)$ is the partial schedule $\sigma^1$ of 
iteration 1. Matrix $T^1$ is similarly represented in the second quarter of Table 2 
(note that the entries in row 0 are updated correspondingly). Now, 
$\delta^2=1$ and hence $\tau^2=4$. The part of the schedule of Fig. 1 corresponding to 
the interval $[0,4)$ is partial schedule $\sigma^2$. Similarly, in the following 
iterations  3 and 4, $\delta^3=1$, $\tau^3=5$, and $\delta^4=1$ and $\tau^4=6$.

Schedule $\sigma^4$ is the resultant complete schedule of Fig. 1 which extends 
through the interval $[0,6)$. Since column 1 is tight, the corresponding parts of job
$J^1$ are included in the decrementing sets in iterations 1,2, 3 and 4 (until this job 
is completely scheduled). Note that $C_{\max}=6$ is attained by both, job $J^1$ and 
machine 2 in the optimal distribution of Fig. 1 (i.e., the total length of job $J^1$ 
and the load of machine 2, see inequalities \ref{2'} and \ref{1'}).
Hence, schedule $\sigma^4$ is optimal. 

Given an optimal distribution of Fig. 1, the extended scheduling procedure will
create an optimal schedule without the knowledge of a solution to the PARTITION
instance. Then the jumps $\delta^i$ will be determined by the lengths of the
selected partition jobs (without our aggregated presentation, the number 
of iterations would depend on $k$, in our case, it would be $k+2$). Note 
that the procedure may split a partition job or/and job $J^2$, which is not
allowed for setting $R |r_j, pmtn \ - \ no split |L_{\max}$. For instance, in 
this example, job $J^2$ was not split on machine 3 the corresponding parts of
that job being selected in consecutive iterations 1,2 and 3 (in fact, we solved 
an instance of problem $R |r_j, pmtn \ - \ no split |L_{\max}$ given a solution 
to the PARTITION instance, see Corollary \ref{sch-con}). Finally, note that,
depending on the made selections of the decrementing sets, the procedure may 
create different optimal schedules respecting the same distribution.\eop

\begin{table}
\centering
\caption{Flowchart of the extended procedure with processing time matrices 
and the corresponding decrementing sets} \label{table2}
\resizebox{17cm}{!}{%
\begin{tabular}{cccccc}

\begin{tabular}{c}
  \\ 
0 \\ 
1 \\ 
2 \\ 
3 \\ 
\end{tabular}
& 
\begin{tabular}{|c|c|c|c|c|}
\hline 
1 & 2 & 3 & 4 & P \\ 
\hline 
0 & 0 & 4 & 0 & 3 \\ 
\hline 
\textcircled{4} & 0 & 2 & 0 & 0 \\ 
\hline 
1 & 0 & 0 & \textcircled{3} & 2 \\ 
\hline 
1 & \textcircled{5} & 0 & 0 & 0 \\ 
\hline 
\end{tabular}
&
\begin{tabular}{|c|c|c|c|c|}
\hline
1 & 2 & 3 & 4 & P\\ 
\hline 
0 & 0 & 1 & 0 & 0\\ 
\hline 
\textcircled{1} & 0 & 2 & 0 & 0\\ 
\hline 
1 & 0 & 0 & 0 & \textcircled{2}\\ 
\hline 
1 & \textcircled{2} & 0 & 0 & 0 \\ 
\hline 
\end{tabular}
&
\begin{tabular}{|c|c|c|c|c|}
\hline
1 & 2 & 3 & 4 & P\\ 
\hline 
0 & 0 & 0 & 0 & 0 \\ 
\hline 
0 & 0 & \textcircled{2} & 0 & 0\\ 
\hline 
\textcircled{1} & 0 & 0 & 0 & 1\\ 
\hline 
1 & \textcircled{1} & 0 & 0 & 0 \\ 
\hline 
\end{tabular}
& 
\begin{tabular}{|c|c|c|c|c|}
\hline
1 & 2 & 3 & 4 & P\\ 
\hline 
0 & 0 & 0 & 0 & 0 \\ 
\hline 
0 & 0 & \textcircled{1} & 0 & 0\\ 
\hline 
0 & 0 & 0 & 0 & \textcircled{3}\\ 
\hline 
\textcircled{1} & 0 & 0 & 0 & 0\\ 
\hline 
\end{tabular}
&
\begin{tabular}{|c|c|c|c|c|}
\hline
1 & 2 & 3 & 4 & P\\ 
\hline 
0 & 0 & 0 & 0 & 0\\ 
\hline 
0 & 0 & 0 & 0 & 0\\ 
\hline 
0 & 0 & 0 & 0 & 0\\ 
\hline 
0 & 0 & 0 & 0 & 0\\ 
\hline 
\end{tabular}
\\
\end{tabular} 
}
\end{table}

\medskip

{\bf Example 1 (continuation 2).} Next, we illustrate the extended schedule 
construction procedure on the problem instance of our basic Example 1. We
first construct an optimal schedule of Fig. 2d respecting optimal distribution 
2 (recall that this schedule is not globally optimal). Table 3 
represents the flowchart of the procedure in its 10 iterations
(we omit the table with all 0 entries of the last iteration 11). Initially
in matrix $T^1$ all entries in row 0 are positive except that of 
column (job) 5. In particular, only row 4 is ready. Since 2 is the minimum
entry in row 0, $\delta^1=2$, hence $\tau^1=2$. The partial schedule of
iteration 1 corresponds to the segment $[0,2)$ of the schedule in Fig. 7.
In the next matrix $T^2$ there arises one additional ready row 2 with the 
entry 5 in column 2 (recall that the minimum entry in column 0 of matrix
$T^1$ was precisely in column 2). Now the minimum entry in row 0 is 3, hence 
$\delta^2=\min \{5,1\}=1$ and $\tau^1=2+1=3$, see again Fig 2d. The entries
in the decrementing set $D^2$ correspond to columns 2 and 5 (both, jobs 2 
and 5 are released by time 2). Again, in the processing time matrix $T^3$
there arises one additional ready row 1 corresponding to job 3. The
decrementing set $D^3$ contains now 3 jobs so that already 3 machines,
1,2 and 4 become busy. At the next iteration 4 job 6 becomes available on
machine 2, but the last unscheduled part of job 2 is scheduled on that machine
by our tie breaking rule. 
The decrementing set of iteration 5 already contains 4 elements, hence all
four machines become busy; $\delta^5=1$ and $\tau^5=2+1+2+2+1=8$, see Fig. 7. 
At iteration 6 all entries in row 0 in  matrix $T^6$ are already 0, hence
all job parts become released. The procedure  continues in the same fashion
until it constructs a complete feasible schedule of Fig. 2d respecting an
optimal distribution 2 to linear program MILP$(C_{\max})$.

\begin{table}
\centering
\caption{caption of table 3.} \label{table3}
\resizebox{18cm}{!}{%
\begin{tabular}{cccccc}
\begin{tabular}{c}
 \\ 
\end{tabular}
& 
\begin{tabular}{ccccc}
\ {\bf 2} \ & \ {\bf 3} \ & \ {\bf 4} \ & \ {\bf 5} \ & \ {\bf 6} \ \\
\end{tabular}
& 
\begin{tabular}{ccccc}
\ {\bf 2} \ & \ {\bf 3} \ & \ {\bf 4} \ & \ {\bf 5} \ & \ {\bf 6} \ \\
\end{tabular}
& 
\begin{tabular}{ccccc}
\ {\bf 2} \ & \ {\bf 3} \ & \ {\bf 4} \ & \ {\bf 5} \ & \ {\bf 6} \ \\
\end{tabular}
& 
\begin{tabular}{ccccc}
\ {\bf 2} \ & \ {\bf 3} \ & \ {\bf 4} \ & \ {\bf 5} \ & \ {\bf 6} \ \\
\end{tabular}
& 
\begin{tabular}{ccccc}
\ {\bf 2} \ & \ {\bf 3} \ & \ {\bf 4} \ & \ {\bf 5} \ & \ {\bf 6} \ \\
\end{tabular}
\\
\\
\begin{tabular}{c}
{\bf 0} \\ 
{\bf 1} \\ 
{\bf 2} \\ 
{\bf 3} \\
{\bf 4} \\
\end{tabular}
& 
\begin{tabular}{|c|c|c|c|c|}
\hline
\ 2 \ & \ 3 \ & \ 8 \ & \ 0 \ & \ 5 \ \\ 
\hline 
0 & 5 & 13 & 0 & 0 \\ 
\hline 
5 & 0 & 0 & 0 & 13 \\ 
\hline 
6 & 10 & 0 & 0 & 0 \\ 
\hline 
5 & 0 & 0 & {\large \textcircled{{\footnotesize 13}}} & 0 \\ 
\hline 
\end{tabular}
&
\begin{tabular}{|c|c|c|c|c|}
\hline
\ 0 \ & \ 1 \ & \ 6 \ & \ 0 \ & \ 3 \ \\ 
\hline 
0 & 5 & 13 & 0 & 0 \\ 
\hline 
{\large \textcircled{{\footnotesize 5}}} & 0 & 0 & 0 & 13 \\ 
\hline 
6 & 10 & 0 & 0 & 0 \\ 
\hline 
5 & 0 & 0 & {\large \textcircled{{\footnotesize 11}}} & 0 \\ 
\hline 
\end{tabular}
&
\begin{tabular}{|c|c|c|c|c|}
\hline
\ 0 \ & \ 0 \ & \ 5 \ & \ 0 \ & \ 2 \ \\ 
\hline 
0 & {\large \textcircled{{\footnotesize 5}}} & 13 & 0 & 0 \\ 
\hline 
{\large \textcircled{{\footnotesize 4}}} & 0 & 0 & 0 & 13 \\ 
\hline 
6 & 10 & 0 & 0 & 0 \\ 
\hline 
5 & 0 & 0 & {\large \textcircled{{\footnotesize 10}}} & 0 \\ 
\hline 
\end{tabular}
&
\begin{tabular}{|c|c|c|c|c|}
\hline
\ 0 \ & \ 0 \ & \ 3 \ & \ 0 \ & \ 0 \ \\ 
\hline 
0 & {\large \textcircled{{\footnotesize 3}}} & 13 & 0 & 0 \\ 
\hline 
{\large \textcircled{{\footnotesize 2}}} & 0 & 0 & 0 & 13 \\ 
\hline 
6 & 10 & 0 & 0 & 0 \\ 
\hline 
5 & 0 & 0 & {\large \textcircled{{\footnotesize 8}}} & 0 \\ 
\hline 
\end{tabular}
&
\begin{tabular}{|c|c|c|c|c|}
\hline
\ 0 \ & \ 0 \ & \ 1 \ & \ 0 \ & \ 0 \ \\ 
\hline 
0 & {\large \textcircled{{\footnotesize 1}}} & 13 & 0 & 0 \\ 
\hline 
0 & 0 & 0 & 0 & {\large \textcircled{{\footnotesize 13}}} \\ 
\hline 
{\large \textcircled{{\footnotesize 6}}} & 10 & 0 & 0 & 0 \\ 
\hline 
5 & 0 & 0 & {\large \textcircled{{\footnotesize 6}}} & 0 \\ 
\hline 
\end{tabular}
\\
\\
\begin{tabular}{c}
{\bf 0} \\ 
{\bf 1} \\ 
{\bf 2} \\ 
{\bf 3} \\
{\bf 4} \\
\end{tabular}
& 
\begin{tabular}{|c|c|c|c|c|}
\hline
\ 0 \ & \ 0 \ & \ 0 \ & \ 0 \ & \ 0 \ \\ 
\hline 
0 & 0 & {\large \textcircled{{\footnotesize 13}}} & 0 & 0 \\ 
\hline 
0 & 0 & 0 & 0 & {\large \textcircled{{\footnotesize 12}}} \\ 
\hline 
{\large \textcircled{{\footnotesize 5}}} & 10 & 0 & 0 & 0 \\ 
\hline 
5 & 0 & 0 & {\large \textcircled{{\footnotesize 5}}} & 0 \\ 
\hline 
\end{tabular}
&
\begin{tabular}{|c|c|c|c|c|}
\hline
\ 0 \ & \ 0 \ & \ 0 \ & \ 0 \ & \ 0 \ \\ 
\hline 
0 & 0 & {\large \textcircled{{\footnotesize 8}}} & 0 & 0 \\ 
\hline 
0 & 0 & 0 & 0 & {\large \textcircled{{\footnotesize 7}}} \\ 
\hline 
0 & {\large \textcircled{{\footnotesize 10}}} & 0 & 0 & 0 \\ 
\hline 
{\large \textcircled{{\footnotesize 5}}} & 0 & 0 & 0 & 0 \\ 
\hline 
\end{tabular}
&
\begin{tabular}{|c|c|c|c|c|}
\hline
\ 0 \ & \ 0 \ & \ 0 \ & \ 0 \ & \ 0 \ \\ 
\hline 
0 & 0 & {\large \textcircled{{\footnotesize 3}}} & 0 & 0 \\ 
\hline 
0 & 0 & 0 & 0 & {\large \textcircled{{\footnotesize 2}}} \\ 
\hline 
0 & {\large \textcircled{{\footnotesize 5}}} & 0 & 0 & 0 \\ 
\hline 
0 & 0 & 0 & 0 & 0 \\ 
\hline 
\end{tabular}
&
\begin{tabular}{|c|c|c|c|c|}
\hline
\ 0 \ & \ 0 \ & \ 0 \ & \ 0 \ & \ 0 \ \\ 
\hline 
0 & 0 & {\large \textcircled{{\footnotesize 1}}} & 0 & 0 \\ 
\hline 
0 & 0 & 0 & 0 & 0 \\ 
\hline 
0 & {\large \textcircled{{\footnotesize 3}}} & 0 & 0 & 0 \\ 
\hline 
0 & 0 & 0 & 0 & 0 \\ 
\hline 
\end{tabular}
&
\begin{tabular}{|c|c|c|c|c|}
\hline
\ 0 \ & \ 0 \ & \ 0 \ & \ 0 \ & \ 0 \ \\ 
\hline 
0 & 0 & 0 & 0 & 0 \\ 
\hline 
0 & 0 & 0 & 0 & 0 \\ 
\hline 
0 & {\large \textcircled{{\footnotesize 2}}} & 0 & 0 & 0 \\ 
\hline 
0 & 0 & 0 & 0 & 0 \\ 
\hline 
\end{tabular}
\\
\end{tabular} 
}
\end{table}


\begin{table}
\centering
\caption{caption of table 4.} \label{table4}
\resizebox{17cm}{!}{%
\begin{tabular}{ccccc}
\begin{tabular}{c}
 \\ 
\end{tabular}
& 
\begin{tabular}{ccccc}
\ {\bf 2} \ & \ {\bf 3} \ & \ {\bf 4} \ & \ {\bf 5} \ & \ {\bf 6} \ \\
\end{tabular}
& 
\begin{tabular}{ccccc}
\ {\bf 2} \ & \ {\bf 3} \ & \ {\bf 4} \ & \ {\bf 5} \ & \ {\bf 6} \ \\
\end{tabular}
& 
\begin{tabular}{ccccc}
\ {\bf 2} \ & \ {\bf 3} \ & \ {\bf 4} \ & \ {\bf 5} \ & \ {\bf 6} \ \\
\end{tabular}
& 
\begin{tabular}{ccccc}
\ {\bf 2} \ & \ {\bf 3} \ & \ {\bf 4} \ & \ {\bf 5} \ & \ {\bf 6} \ \\
\end{tabular}
\\
\\
\begin{tabular}{c}
{\bf 0} \\ 
{\bf 1} \\ 
{\bf 2} \\ 
{\bf 3} \\
{\bf 4} \\
\end{tabular}
& 
\begin{tabular}{|c|c|c|c|c|}
\hline
\ 2 \ & \ 3 \ & \ 8 \ & \ 0 \ & \ 5 \ \\ 
\hline 
0 & 5 & 13 & 0 & 0 \\ 
\hline 
5 & 0 & 0 & 0 & 13 \\ 
\hline 
4 & 10 & 0 & 0 & 0 \\ 
\hline 
7 & 0 & 0 & {\large \textcircled{{\footnotesize 13}}} & 0 \\ 
\hline 
\end{tabular}
&
\begin{tabular}{|c|c|c|c|c|}
\hline
\ 0 \ & \ 1 \ & \ 6 \ & \ 0 \ & \ 3 \ \\ 
\hline 
0 & 5 & 13 & 0 & 0 \\ 
\hline 
{\large \textcircled{{\footnotesize 5}}} & 0 & 0 & 0 & 13 \\ 
\hline 
4 & 10 & 0 & 0 & 0 \\ 
\hline 
7 & 0 & 0 & {\large \textcircled{{\footnotesize 11}}} & 0 \\ 
\hline 
\end{tabular}
&
\begin{tabular}{|c|c|c|c|c|}
\hline
\ 0 \ & \ 0 \ & \ 5 \ & \ 0 \ & \ 2 \ \\ 
\hline 
0 & {\large \textcircled{{\footnotesize 5}}} & 13 & 0 & 0 \\ 
\hline 
{\large \textcircled{{\footnotesize 4}}} & 0 & 0 & 0 & 13 \\ 
\hline 
4 & 10 & 0 & 0 & 0 \\ 
\hline 
7 & 0 & 0 & {\large \textcircled{{\footnotesize 10}}} & 0 \\ 
\hline 
\end{tabular}
&
\begin{tabular}{|c|c|c|c|c|}
\hline
\ 0 \ & \ 0 \ & \ 1 \ & \ 0 \ & \ 0 \ \\ 
\hline 
0 & {\large \textcircled{{\footnotesize 1}}} & 13 & 0 & 0 \\ 
\hline 
0 & 0 & 0 & 0 & {\large \textcircled{{\footnotesize 13}}} \\ 
\hline 
{\large \textcircled{{\footnotesize 4}}} & 10 & 0 & 0 & 0 \\ 
\hline 
7 & 0 & 0 & {\large \textcircled{{\footnotesize 6}}} & 0 \\ 
\hline 
\end{tabular}
\\
\\
\begin{tabular}{c}
{\bf 0} \\ 
{\bf 1} \\ 
{\bf 2} \\ 
{\bf 3} \\
{\bf 4} \\
\end{tabular}
& 
\begin{tabular}{|c|c|c|c|c|}
\hline
\ 0 \ & \ 0 \ & \ 0 \ & \ 0 \ & \ 0 \ \\ 
\hline 
0 & 0 & {\large \textcircled{{\footnotesize 13}}} & 0 & 0 \\ 
\hline 
0 & 0 & 0 & 0 & {\large \textcircled{{\footnotesize 12}}} \\ 
\hline 
{\large \textcircled{{\footnotesize 3}}} & 10 & 0 & 0 & 0 \\ 
\hline 
7 & 0 & 0 & {\large \textcircled{{\footnotesize 5}}} & 0 \\ 
\hline 
\end{tabular}
&
\begin{tabular}{|c|c|c|c|c|}
\hline
\ 0 \ & \ 0 \ & \ 0 \ & \ 0 \ & \ 0 \ \\ 
\hline 
0 & 0 & {\large \textcircled{{\footnotesize 9}}} & 0 & 0 \\ 
\hline 
0 & 0 & 0 & 0 & {\large \textcircled{{\footnotesize 8}}} \\ 
\hline 
0 & {\large \textcircled{{\footnotesize 10}}} & 0 & 0 & 0 \\ 
\hline 
7 & 0 & 0 & {\large \textcircled{{\footnotesize 1}}} & 0 \\ 
\hline 
\end{tabular}
&
\begin{tabular}{|c|c|c|c|c|}
\hline
\ 0 \ & \ 0 \ & \ 0 \ & \ 0 \ & \ 0 \ \\ 
\hline 
0 & 0 & {\large \textcircled{{\footnotesize 8}}} & 0 & 0 \\ 
\hline 
0 & 0 & 0 & 0 & {\large \textcircled{{\footnotesize 7}}} \\ 
\hline 
0 & {\large \textcircled{{\footnotesize 8}}} & 0 & 0 & 0 \\ 
\hline 
{\large \textcircled{{\footnotesize 7}}} & 0 & 0 & 0 & 0 \\ 
\hline 
\end{tabular}
&
\begin{tabular}{|c|c|c|c|c|}
\hline
\ 0 \ & \ 0 \ & \ 0 \ & \ 0 \ & \ 0 \ \\ 
\hline 
0 & 0 & {\large \textcircled{{\footnotesize 1}}} & 0 & 0 \\ 
\hline 
0 & 0 & 0 & 0 & 0 \\ 
\hline 
0 & {\large \textcircled{{\footnotesize 1}}} & 0 & 0 & 0 \\ 
\hline 
0 & 0 & 0 & 0 & 0 \\ 
\hline 
\end{tabular}
\\
\end{tabular} 
}
\end{table}

Now we apply our schedule construction procedure to a non-optimal
distribution to obtain a globally optimal schedule respecting that 
distribution. We illustrate this using a non-optimal distribution 3 from 
Example 1 (see Fig. 2e). We represent the flowchart of the procedure in Table 4. 
Initially at iteration 1  we have one ready row 4 with a single valid entry 13
(corresponding to job 2) which is included into the decrementing set $D^1$; 
$\delta_1=t^1_{02}=2$ and $\tau^1=2$. At iteration 2 row 2 also gets ready with 
a single valid entry 5 corresponding to job 2; $D^2= \{ t^2_{22}, t^2_{45}\} 
=\{5,11\}$, $\delta_2=t^2_{03}=1$ and $\tau^2=2+1=3$. At iteration 3 one additional
row 1  with a single valid entry 5 corresponding to job 2 gets ready; 
$D^3= \{ t^3_{13}, t^3_{22}, t^3_{45}\} =\{5,4,10\}$; now  $\delta_3=t^3_{06}=2$, 
but since the remaining processing time of job 2 is 13, the same as that of 
job 6, by our tie breaking rule we include $t^3_{22}$ (and not $t^3_{26}$) in set 
$D^3$ (without creating unnecessary preemption of job 2). We respectively skip
one iteration by setting $\delta_3=4$ and hence letting $\tau^3=2+1+4=7$. At
iteration 4 all 4 rows are ready and the corresponding four entries marked in the
fourth quarter of Table 4 are included in set $D^4$. The next four iterations 
are similarly reflected in Table 4 (we omitted gain the 0-matrix of iteration 9). 
The resultant complete optimal schedule with makespan 
$\tau^8=2+1+4+1+3+2+7+1=21$ coincides with that of Fig. 2e.\eop

\section{Concluding remarks}

Optimal distributions to linear programs that we considered here,
including linear program MILP$(C_{\max})$, do not ``completely capture'' all 
required features for the construction of an optimal schedule to problem 
$R |r_j; pmtn |C_{\max}$. Some optimal distributions may posses better 
features than others for the creation of an optimal schedule, 
but it is difficult to predict what kind of an optimal distribution 
an LP solver will deliver and what kind of an optimal distribution is
preferable, in general. Moreover, a non-optimal distribution may suit 
better a given problem instance than any optimal one so that an optimal
schedule respecting that non-optimal distribution may be globally optimal, 
whereas no globally optimal schedule respecting an optimal distribution may 
exist (Section 6).  

The extended schedule construction procedure from Section 7.1 may be applied
to any (not necessarily optimal) distribution to obtain an optimal schedule
respecting that distribution. Such a schedule may also be globally 
optimal (e.g., the schedule of Fig. 4d). The method of Lawler and Labetoulle 
\cite{ll78} for the construction of an optimal schedule relies on 
the fact that the makespan of a (globally) optimal schedule
to linear program LP2$(C_{\max })$ equals to the corresponding $C_{\max}$ 
for problem $R |pmtn |C_{\max}$. As we saw, similar property does not hold 
for problem $R |r_j; pmtn |C_{\max}$, as the makespan of a globally
optimal schedule may be larger than the corresponding $C_{\max}$. 

A linear program, ``more intelligent'' than the linear programs studied here, 
would ``correctly'' estimate the starting time of the first scheduled job on 
each machine. However, this kind of estimation looks unrealistic  without 
actually carrying out the scheduling of assigned job parts. 
Hence unlikely, there exists a liner programming such that an 
optimal schedule respecting an optimal distribution to that linear program
can be guaranteed to be optimal. Because of this, we suggested an alternative 
schedule construction procedure that delivers an optimal schedule to problem 
$R |r_j, pmtn |L_{\max}$ respecting any distribution. Such distribution can
of course be obtained by a linear program, but it might be possible to create
it using some alternative way. In contrast with the studied here scheduling problem 
$R |r_j, pmtn |L_{\max}$, for the setting $R |pmtn |L_{\max}$ without job
release times, an optimal schedule respecting an optimal distribution to
linear program LP2$(C_{\max })$ is guaranteed to be optimal. In this sense, the 
latter problem possesses more accessible structural properties than the former one.

\section{Acknowledgment}

The author is grateful to Victor Pacheco for his very kind help in 
creating figures and tables using the tools of a graphical interface
Xfig and Latex, respectively.

\end{document}